\documentclass[a4paper,10pt]{article}
\usepackage[margin=2cm]{geometry}
\usepackage[utf8]{inputenc} 
\usepackage{amsmath}
\usepackage{multicol}
\usepackage{graphicx}
\usepackage[dvipsnames]{xcolor}
\usepackage{float}
\usepackage{xltabular}
\usepackage{booktabs}
\usepackage{array}
 \usepackage{amssymb}
\usepackage{caption}
\usepackage{multirow}
\usepackage{amsfonts}
\usepackage{adjustbox}
\usepackage{caption}
\usepackage{subcaption}
\usepackage{epstopdf}
\usepackage{orcidlink}
\usepackage[style=chem-angew,doi=false,url=false,isbn=false]{biblatex}

\usepackage[nameinlink,capitalize]{cleveref}
\newbibmacro{string+doiurl}[1]{%
  \iffieldundef{doi}
    {\iffieldundef{url}
       {#1}
       {\href{\thefield{url}}{#1}}}
    {\href{https://doi.org/\thefield{doi}}{#1}}}

\makeatletter
\def\blx@driver#1{%
  \ifcsdef{blx@bbx@#1}
    {\usebibmacro{string+doiurl}{\csuse{blx@bbx@#1}}}
    {\ifcsdef{blx@bbx@*}
       {\blx@warning{%
          No driver for entry type '#1'.\MessageBreak
          Using fallback driver}%
        \usebibmacro{string+doiurl}{\csuse{blx@bbx@*}}}
       {\blx@error
          {No driver found}
          {I can't find a driver for the entry type
           '\abx@field@entrytype'\MessageBreak
           and there is no fallback driver either}}}}
\makeatother

\hypersetup{colorlinks,
linkcolor={blue},
citecolor={green!60!black},
urlcolor={cyan}
}

\title{\bf Possible Formation of Traversable Wormholes and Their Thermodynamic Analysis in $\mathcal{F}(Q,\mathcal{L}_{m},\mathcal{T})$ Gravity}

\author{\bf{Rounak Manna$^1$}  \orcidlink{0009-0002-9373-6414} \thanks {mannarounak97@gmail.com}~,~~ \bf{Krishna Pada Das $^1$} \orcidlink{0000-0003-4653-6006} \thanks {krishnapada0019@gmail.com}~~and~  
\bf{Ujjal Debnath$^1$}  \orcidlink{0000-0002-2124-8908} \thanks{ujjaldebnath@gmail.com}\\
$^1$Department of Mathematics, Indian Institute of Engineering\\
Science and Technology, Shibpur, Howrah-711 103, India.\\
}
\date{\today}
\begin{document}
\maketitle

\begin{abstract}
In this work, we investigate static and spherically symmetric traversable wormhole solutions within the framework of the extended symmetric teleparallel gravity, specifically the $\mathcal{F}(Q,\mathcal{L}_{m},\mathcal{T})$ gravity theory, where $Q$, $\mathcal{L}_{m}$, and $\mathcal{T}$ are the respective representations of the non-metricity scalar, the matter Lagrangian, and the trace of the energy-momentum tensor. By employing a specific redshift function and deriving the shape function through the Karmarkar condition, we examine the fundamental geometric features required for a viable wormhole structure. The analysis confirms the satisfaction of key conditions such as the throat condition, flaring-out condition, and asymptotic flatness. A detailed study of energy conditions for various values of model parameters reveals that the null energy condition and averaged null energy condition are violated near the throat, indicating the presence of exotic matter. Additionally, thermodynamic quantities such as temperature, pressure, specific heat, work density, and energy flux are analyzed, all of which support the thermal and equilibrium stability of the wormhole. Our findings demonstrate that even in extended theories like $\mathcal{F}(Q,\mathcal{L}_{m},\mathcal{T})$ gravity, exotic matter remains essential for sustaining traversable wormholes. This work lays the foundation for further investigations into their stability under dynamical perturbations and potential astrophysical implications.\\

\end{abstract}

\section{Introduction}
To date, gravitational theories, including the foundational theory of General Relativity (GR), have been based on the framework of Riemannian geometry. This geometric structure employs a Levi-Civita connection that is both symmetric and metric-compatible, with a non-vanishing curvature. In GR, the fundamental dynamical variables are the metric tensor and its derivatives. The theory assumes a torsion-free connection, specifically the Levi-Civita connection. In contrast, to incorporate torsion into the geometric framework of gravity, Einstein introduced another representation identified as the Teleparallel Equivalent of General Relativity (TEGR) \cite{einstein1928}. This approach is founded on a different geometrical framework that uses the Weitzenböck connection \cite{Weitzenbock1923,cho1976einstein,obukhov2006covariance,hammond2002torsion,aldrovandi2012teleparallel}. Unlike the Levi-Civita connection, the Weitzenböck connection has vanishing Riemann curvature $R$ but non-zero torsion $T$, thus providing a distinct geometric description of gravitation. An alternative framework of interest is Symmetric Teleparallel Gravity (STG) \cite{nester1998symmetric}, which is the primary focus of our study. This theory is built upon a connection that is both curvature-free and torsion-free, referred to as the symmetric teleparallel connection, commonly referred to as the Weyl–Cartan connection, and that does not preserve metric compatibility. Within this connection, the teleparallel condition ensures vanishing curvature, while symmetry guarantees the absence of torsion. The geometric structure underlying STG is characterized by the non-metricity term $Q$ of spacetime, which encapsulates its non-trivial gravitational features. \\

GR continues to be the most successful and precisely tested theory of gravity, with strong experimental validation from Earth-based and Solar System observations \cite{will2014confrontation}. However, from a theoretical standpoint, it presents significant challenges like its non-renormalizability \cite{addazi2022quantum}, which prevents its integration into a consistent quantum framework \cite{batista2025white}. Moreover, on cosmological scales, unresolved phenomena such as dark matter and dark energy remain open questions \cite{copeland2006dynamics,cai2010quintom}. Thus, a modification of GR appears necessary, either in its geometric (curvature) formulation or in its treatment of the matter (energy–momentum) sector. Over the past decades, numerous efforts have been made to extend GR within various theoretical frameworks. The best approach involves starting with the Einstein-Hilbert Lagrangian and introducing additional terms, leading to modified gravity theories such as  $\mathcal{F}(R)$ gravity \cite{sotiriou2010f,nojiri2011unified,capozziello2011extended}, $\mathcal{F}(G)$ gravity \cite{nojiri2005modified,bamba2017energy}, Lovelock gravity \cite{lovelock1971einstein}, and others. In addition, one may explore geometrical frameworks other than Riemannian geometry, such as torsion-based formulations of gravity. In this connection, some extensions of TEGR such as $\mathcal{F} (T)$ gravity \cite{cai2016f}, $\mathcal{F}(T,\mathcal{T})$ gravity \cite{harko2014f}, $\mathcal{F}(T,B)$ gravity \cite{bahamonde2015modified}, $\mathcal{F}(T,T_{G})$ gravity \cite{kofinas2014teleparallel}, and $\mathcal{F}(T,B,\mathcal{T})$ gravity \cite{das2024anisotropic}. Along similar lines, one can instead consider non-metricity, which leads to symmetric teleparallel gravity \cite{jimenez2018coincident}, $\mathcal{F}(Q)$ gravity \cite{jimenez2018coincident,anagnostopoulos2021first}, etc.\\

Now, the $\mathcal{F}(Q,\mathcal{T})$ gravity framework arises as an advancement of the $\mathcal{F}(Q)$ theory, characterized by a non-minimal coupling between the non-metricity scalar $Q$ and the trace $\mathcal{T}$ of the energy-momentum tensor \cite{xu2019f}. Consequently, in ref.~\cite{xu2020weyl}, authors have explored such extended gravitational theory in the framework of the proper Weyl geometry. Later, in refs.~\cite{arora2020energy,arora2020f,pati2023evolutionary,el2023constant,najera2022cosmological,godani2021frw}, authors have shown different types of cosmological applications. Also, in ref.~\cite{tayde2023wormhole}, authors have constructed wormhole solutions through the embedding procedure, and spherically symmetric stellar structure has been explored in $\mathcal{F}(Q,\mathcal{T})$ gravity \cite{das2024spherically}. Recently, by incorporation of boundary term $B$ in the Lagrangian density, another extension, so-called $\mathcal{F}(Q, B)$ gravity, has been formulated in \cite{de2024non,lohakare2024stability}. Here, the term $B$ actually comes from the standard Levi-Civita Ricci scalar that is $\mathring{R} = Q + B$. Inspired by the theory of non-minimal coupling upon the $\mathcal{F}(Q)$ gravitational framework, in the literature \cite{hazarika2024f}, authors have introduced a new type of extension that is $\mathcal{F}(Q,\mathcal{L}_{m})$ gravity (as similar mechanism of $\mathcal{F}(R,\mathcal{L}_{m})$ theory \cite{harko2010f}), in which the matter Lagrangian $\mathcal{L}_{m}$ is incorporated into the Lagrangian density of $\mathcal{F}(Q)$ gravity. This framework allows for the inclusion of both minimal and non-minimal interactions between matter and geometry, offering a versatile framework to investigate their mutual interaction within the fabric of spacetime.  Subsequently, a variety of cosmological models have been explored against the background of $\mathcal{F}(Q,\mathcal{L}_{m})$ gravity \cite{myrzakulov2024late,myrzakulov2024observational,myrzakulov2025modified}. During this exploration, we will focus on a further extension of $\mathcal{F}(Q,\mathcal{L}_{m})$ gravity with the incorporation of the scalar term $\mathcal{T}$ in the Lagrangian density of $\mathcal{F}(Q,\mathcal{L}_{m})$ gravity, that is, $\mathcal{F}(Q,\mathcal{L}_{m},\mathcal{T})$ gravity, as in a similar approach to $\mathcal{F}(R,\mathcal{T})$, $\mathcal{F}(T, \mathcal{T})$ and $\mathcal{F}(Q,\mathcal{T})$ gravity. The main motivation comes from the literature \cite{mota2024neutron}, in which the authors have extended $\mathcal{F}(R,\mathcal{L}_{m})$ gravity as $\mathcal{F}(R,\mathcal{L}_{m},\mathcal{T})$ gravity and also, several models \cite{de2025palatini,errehymy2025einstein,pretel2024moment,tangphati2025anisotropic} have been developed in such gravitational theory.\\

Interstellar travel has attracted considerable scientific interest and possesses a rich historical background. However, its practical feasibility remains an unresolved question in the domain of gravitational physics. The idea of a conduit connecting distant regions of spacetime was initially speculative until the formulation of GR in 1915 provided a theoretical framework for such possibilities. Within this context, the notion of a \emph{wormhole} emerged as hypothetical structures resulting from non-trivial spacetime topology that could serve as shortcuts between two separate spacetimes or distant regions within the same spacetime. Among various compact relativistic configurations, wormholes stand out as particularly fascinating theoretical constructs: they are mathematically consistent solutions of Einstein's field equations, yet they remain unsupported by direct observational evidence. Physically, a wormhole is often envisioned as a tunnel-like geometry that bridges distinct spacetime points, potentially offering a means of rapid traversal across vast cosmic distances. Historically, the notion referred to as a wormhole was first labeled as such by Wheeler~\cite{wheeler1957nature}, who envisioned them as entities emerging from the spacetime quantum foam, connecting distinct regions of spacetime at the Planck scale. However, as insightfully reviewed in~\cite{fuller1962causality}, these early solutions were non-traversable and collapsed almost instantaneously upon their formation. Subsequently, the groundbreaking research by Morris and Thorne~\cite{MorrisThorne1988} initiated the study of static wormholes. They considered static, spherically symmetric spacetime geometries and explored the possibility of traversable wormholes, hypothetical astrophysical objects maintained by a single fluid that violates the null energy condition (NEC).\\

Currently, considerable attention is being devoted to the investigation of traversable wormholes that are free from singularities and horizons. It is well established that the construction of a wormhole using ordinary matter, that is, matter which satisfies the standard energy conditions, poses a significant challenge in gravitational physics. In this regard, it is predicted that, within modified gravity frameworks, it is possible to form wormholes that do not necessitate exotic matter, at least in the region surrounding the throat. Therefore, the formation of a traversable wormhole typically necessitates the violation of certain energy conditions, which implies the existence of exotic matter in the vicinity of the wormhole throat. To date, numerous studies have been conducted on the physical properties and exact solutions of various types of wormholes. Notable developments include static wormholes~\cite{IdaHayward1999,FewsterRoman2005,Kuhfittig2006,Rahaman2006,JamilKuhfittigRahamanRakib2010}, dynamical wormholes~\cite{HochbergVisser1998,Hayward1999}, and rotating wormholes~\cite{Teo1998,Kuhfittig2003}, proposed by several authors. We now turn our attention to the investigation of wormhole solutions within the framework of modified theories of gravity, with particular emphasis on the symmetric teleparallel framework and its extensions. In a series of articles~\cite{banerjee2021wormhole,sharma2022traversable,mustafa2021wormhole,mustafa2022traversable,hassan2022embedding,parsaei2022wormhole,kiroriwal2023new,kiroriwal2024wormhole}, authors have investigated wormhole solutions in the framework $\mathcal{F}(Q)$ and $\mathcal{F}(Q,\mathcal{T})$ gravities. \\

The deep connection between gravity and thermodynamics was first revealed through Hawking’s discovery that black holes emit radiation with a temperature proportional to surface gravity and an entropy proportional to the area of the event horizon \cite{Hawking:1975vcx, Hawking:1976de, Gibbons:1977mu}. This insight suggested that gravity might have a thermodynamic origin. Building on this idea, Jacobson \cite{Jacobson:1995ab} later demonstrated that Einstein’s field equations could be derived from the fundamental principles of thermodynamics. Since then, many researchers have investigated the thermodynamic properties of black holes \cite{10.1063/1.2913906, Wald:1979zz, Ong:2022frf, Aman:2005xk, Witten:2024upt}, including temperature, entropy, and energy flux, to better understand the structure of spacetime. Motivated by these developments, similar studies have been extended to wormholes, which, though horizonless, may display thermodynamic features similar to black holes \cite{Hayward:1998pp}. In this context, investigating the thermodynamic properties of wormholes within the framework of extended theories of gravity, including non-Riemannian geometry, provides a compelling approach to understanding their thermal behavior and energy characteristics near the throat. Several notable studies have focused on the thermodynamic aspects of wormholes, exploring their thermal behavior, stability, and energy characteristics from various theoretical perspectives \cite{Hong:2003dj, keathley2022detectability, martin2011lorentzian, Ditta:2025wwx, Saiedi:2012qk, Debnath:2012fqp}.\\

In the current work, our primary objective is to investigate the possible existence of traversable wormhole solutions and its thermodynamical analysis in the context of more extended symmetric teleparallel gravity i.e., $\mathcal{F}(Q,\mathcal{L}_{m},\mathcal{T})$ gravity. In this investigation, we employ the embedding approach to obtain traversable wormhole solutions. The embedding method facilitates the incorporation of the nontrivial effects arising from $\mathcal{F}(Q,\mathcal{L}_{m},\mathcal{T})$ gravity into the effective equations governing the energy density and pressures.  An important advantage of this approach lies in the ability to analyze the energy conditions by employing the standard constraints derived from the Raychaudhuri equation. The embedding framework further facilitates a systematic examination of the null, dominant, and strong energy conditions in the context of wormhole geometries. In this context, readers may refer to the literature \cite{hassan2022embedding,mustafa2024new, Das:2024qmi,naz2023evolving,chaudhary2024stability,paramanik2025embedding, Manna:2025ecw}, where the authors have investigated wormhole solutions using the embedding procedure.\\
The subsequent content of this paper is laid out as follows: Following this introductory discussion, we present a concise overview of the mathematical formulation of an extended gravity theory, namely $\mathcal{F}(Q,\mathcal{L}_{m},\mathcal{T})$ gravity, which incorporates the trace of the energy-momentum tensor as an additional geometrical contribution. In section~\ref{sec3}, we report independent field equations corresponding to static, spherically symmetric Morris-Thorne type wormhole geometry and solve to construct thermodynamical parameters. Next, in section~\ref{sec4}, the wormhole geometry has been obtained using the embedding class-1 technique. Further, in two consecutive sections like \ref{sec5} and \ref{sec6}, we examine the necessary conditions of a traversable wormhole through embedding diagrams (two and three-dimensional) and proper radial distance, respectively. In section~\ref{sec7}, after briefly introducing energy conditions, we have examined the possible formation of a traversable wormhole through the violation of these conditions in the vicinity of the throat region. In addition, we have incorporated an analysis concerning the presence of exotic matter near the throat, associated with the violation of the averaged null energy condition (ANEC) in section \ref{7.2}. The thermodynamic analysis of traversable wormholes in the context of our proposed gravity has been analyzed in section~\ref{sec8}. Finally, in section~\ref{sec10}, we provide a comprehensive summary of our study along with the main conclusions. Throughout this work, we consistently employ the spacetime metric signature $(+,-,-,-)$ and adopt geometrized units by setting the universal constants like Newton's gravitational constant $G$ and the speed of light in vacuum $c$, equal to unity, i.e. $G = c = 1$.

%===========================================================================================================
\section{Mathematical structure of $\mathcal{F}(Q,\mathcal{L}_{m},\mathcal{T})$ gravity}\label{sec2}

In the context of differential geometry, a general expression for a linear affine connection can be formulated as follows \cite{hehl1995metric}
\begin{eqnarray}\label{1}
\Gamma^{\xi}_{~\mu\nu}={\big\{}^{\xi}_{~\mu\nu}{\big\}} + K^{\xi}_{~\mu\nu} + L^{\xi}_{~\mu\nu}~,
\end{eqnarray}
where ${\big\{}^{\xi}_{~\mu\nu}{\big\}}$ denotes the Christoffel symbols, which serve as the components of the Levi-Civita connection compatible with the metric $g_{\mu\nu}$, and is defined as
\begin{eqnarray}\label{2}
{\big\{}^{\xi}_{~\mu\nu}{\big\}}=\frac{1}{2}g^{\xi\chi}\big(\partial_{\mu}g_{\chi\nu} + \partial_{\nu}g_{\chi\mu} - \partial_{\chi}g_{\mu\nu}\big)~.
\end{eqnarray}
The contortion tensor is
\begin{eqnarray}\label{3}
K^{\xi}_{~\mu\nu}=\frac{1}{2}T^{\xi}_{~\mu\nu} + T_{(\mu~\nu)}^{~~\xi}~,
\end{eqnarray}
with the torsion tensor $T^{\xi}_{~\mu\nu}\equiv
2\Gamma^{\xi}_{~\mu\nu}$ and the deformation tensor can be given in terms of the non-metricity tensor as
\begin{eqnarray}\label{4}
L^{\xi}_{~\mu\nu}=\frac{1}{2}g^{\xi\chi}\big(-Q_{\mu\chi\nu} - Q_{\nu\chi\mu} + Q_{\chi\mu\nu}\big)~.
\end{eqnarray}
In this framework, the non-metricity tensor $Q_{\gamma\mu\nu}$ is described as $Q_{\gamma\mu\nu} \equiv \nabla_{\gamma} g_{\mu\nu}$, representing the covariant derivative of the metric tensor with respect to the Weyl-Cartan connection $\Gamma^{\xi}_{\mu\nu}$. This quantity can be explicitly expressed as follows.
\begin{eqnarray}\label{5}
Q_{\gamma\mu\nu}=-\frac{\partial g_{\mu\nu}}{\partial x^{\gamma}} + g_{\nu\xi}\Gamma^{\xi}_{~\mu\gamma} + g_{\xi\mu}\Gamma^{\xi}_{~\nu\gamma}~.
\end{eqnarray}
Accordingly, the non-metricity quantity $Q$ takes the form
\begin{eqnarray}\label{6}
Q \equiv -g^{\mu\nu}\Big(L^{\xi}_{~\chi \mu}L^{\chi}_{~\nu\xi} - L^{\xi}_{~\chi\xi}L^{\chi}_{~\mu\nu}\Big)
\end{eqnarray}
with
\begin{eqnarray}\label{7}
L^{\xi}_{~\chi\gamma} \equiv -\frac{1}{2}g^{\xi \lambda}\Big(\bigtriangledown_{\gamma} g_{\chi\lambda} + \bigtriangledown_{\chi} g_{\lambda\gamma}- \bigtriangledown_{\lambda} g_{\chi\gamma}\Big)
\end{eqnarray}
\begin{eqnarray}\label{8}
Q_{\xi\mu\nu}=\nabla_{\xi}g_{\mu\nu}.
\end{eqnarray}
The non-metricity tensor admits the following independent trace components.
\begin{eqnarray}\label{9}
 Q_{\xi} \equiv Q_{\xi}^{~\mu}{_\mu}~~~~~\text{and}~~~~ \tilde{Q}_{\xi} \equiv Q^{\mu}_{~\xi\mu}~.
\end{eqnarray}
The non-metricity tensor also allows for the definition of the superpotential function, which is written as
\begin{eqnarray}\label{10}
P^{\xi}_{~\mu\nu}=\frac{1}{4}\Big[-Q^{\xi}_{~\mu\nu} + 2Q_{(\mu~\nu)}^{~~\xi} + Q^{\xi}g_{\mu\nu} - \tilde{Q}^{\xi}g_{\mu\nu} -\delta^{\xi}{_(\mu  Q_{\nu)}}\Big]~.
\end{eqnarray}
The scalar quantity $Q$, interpreted as the trace of the non-metricity tensor, can be written in the following expression.
\begin{eqnarray}\label{11}
Q=-Q_{\xi\mu\nu}P^{\xi\mu\nu}~.
\end{eqnarray}

Within the theoretical setting of $\mathcal{F}(Q,\mathcal{L}_{m},\mathcal{T})$ gravity, the gravitational action is expressed as: 
\begin{equation}\label{12}
\mathcal{S}=\frac{1}{16\pi}\int \mathcal{F}(Q,\mathcal{L}_{m},\mathcal{T})\sqrt{-g}d^{4}x + \int \mathcal{L}_{m}\sqrt{-g}d^{4}x
\end{equation}
Here, $\mathcal{F}(Q,\mathcal{L}_{m},\mathcal{T})$ is an arbitrary function of the non-metricity term $Q$ matter Lagrangian $\mathcal{L}_{m}$, and the trace of the energy-momentum tensor $\mathcal{T}$. In addition, $g$ is the determinant of the metric tensor $g_{\mu\nu}$, that is, $g=det\left(g_{\mu\nu}\right)$. \\
Varying the action \cref{12}, we can get the following field equation
\begin{multline}\label{13}
\frac{2}{\sqrt{-g}}\nabla_{\zeta}\left(\sqrt{-g}\mathcal{F}_{Q}P^{\zeta}_{~\mu\nu}\right) + \frac{1}{2}g_{\mu\nu}\mathcal{F} + \mathcal{F}_{Q}\left(P_{\mu\zeta\chi}Q_{\nu}^{~\zeta\chi} - 2 Q_{\zeta\chi\mu}P^{\zeta\chi}_{~~\nu}\right) = -8\pi \mathcal{T}_{\mu\nu} + \frac{1}{2}\mathcal{F}_{\mathcal{L}_{m}}\left(\mathcal{L}_{m}g_{\mu\nu} - \mathcal{T}_{\mu\nu}\right)\\ + \mathcal{F}_{\mathcal{T}}\left(\mathcal{T}_{\mu\nu} + \Theta_{\mu\nu}\right),
\end{multline}
where the energy–momentum tensor is defined as
\begin{eqnarray}\label{14}
\mathcal{T}_{\mu\nu}=-\frac{2}{\sqrt{-g}}\frac{\delta\left(\sqrt{-g}\mathcal{L}_{m}\right)}{\delta g^{\mu\nu}}=g_{\mu\nu}\mathcal{L}_{m}-2\frac{\partial \mathcal{L}_{m}}{\partial g^{\mu\nu}},
\end{eqnarray}
$\text{with the trace}~~~\mathcal{T}=g^{\mu\nu}\mathcal{T}_{\mu\nu},~\text{and the tensor}~~~\Theta_{\mu\nu}=g^{\zeta\chi}\frac{\delta\mathcal{T}_{\zeta\chi}}{\delta g^{\mu\nu}}$.
%====================================================================================================
\section{Wormhole spacetime and the field equations}\label{sec3}

Consider a static, spherically symmetric line element to represent wormhole geometry \cite{MorrisThorne1988} as follows:
\begin{equation}\label{me15}
ds^{2}=e^{\Phi(r)}dt^{2} - \frac{dr^{2}}{\left[1-\frac{b(r)}{r}\right]} - r^{2}\left(d\theta^{2} + \sin^{2}\theta d\phi^{2}\right) ,
\end{equation}
where the functions $\Phi(r)$ and $b(r)$ contain essential details about the geometry of the wormhole. $b(r)$ is referred to as the shape function, while $\Phi(r)$ is called the redshift function. In essence, the shape function explains the way in which spatial dimensions alter as one passes through the wormhole, and the redshift function sheds light on the gravitational pull at various radial positions. It measures the extent to which the gravitational field compresses or stretches light or communications coming from a faraway source.

In order to acquire a traversable wormhole, the shape and redshift functions need to meet the following minimal specifications \cite{Morris:1988cz}:
\begin{enumerate}
    \item The event horizon ought to be eliminated; specifically, $\Phi(r)$ must be both finite and non-zero at the throat.\label{cond1}
    \item At the wormhole's throat, that is, $r=r_t$, $b(r)$ must meet the throat requirement specified by $b(r_t)=r_t$ and $1-\frac{b(r)}{r}>0$ for all $r\in(r_t,\infty)$.\label{cond2}
    \item In order to satisfy the flaring-out requirement, $b(r)$ needs to match the outcomes of the embedding computation, which is described by $\ \frac {b(r)-r b '(r)}{b^2 (r)}>0$ for all $r\in(r_t,\infty)$.\label{cond3}
    \item The prior criterion can be simplified as follows at the wormhole throat: $b '(r_t)<1$.\label{cond4}
    \item Furthermore, as $r\to \infty$, the ratio $\frac{b(r)}{r}$ must approach 0 according to the asymptotic flatness requirement.\label{cond5}
\end{enumerate}

\noindent{Now,} Corresponding to the metric \cref{me15}, the expression of the non-metricity term $Q$ can be obtained as 
\begin{eqnarray}\label{17}
Q=-\frac{b}{r^{2}}\left[\frac{rb' - b}{r(r - b)} + \Phi'\right].
\end{eqnarray}
The energy–momentum tensor characterizing an anisotropic fluid distribution can be expressed as follows \cite{das2024anisotropic}:
\begin{eqnarray}\label{16}
\mathcal{T}_{\mu\nu}=\left(\rho + p_{r}\right)U_{\mu}U_{\nu} - p_{t}g_{\mu\nu} + \left(p_{r} - p_{t}\right)V_{\mu}V_{\nu},
\end{eqnarray}
where $U^\mu$ and $V^\mu$ denote the four-velocity and four-radial vectors, respectively, while the quantities $\rho$, $p_r$, and $p_t$ represent the energy density, the radial pressure, and the tangential pressure of the fluid. Thus, the trace of the energy–momentum tensor is given as
\begin{eqnarray}
\mathcal{T}=\rho-p_{r}-2p_{t}.
\end{eqnarray}
To simplify the analysis, we consider a particular functional form given by
\begin{eqnarray}\label{18}
\mathcal{F}(Q,\mathcal{L}_{m},\mathcal{T})=f(Q) + \alpha \mathcal{L}_{m} + \beta \mathcal{T},
\end{eqnarray}
where $\alpha$ and $\beta$ are constant parameters.
Without loss of generality, we consider $\mathcal{L}_{m}=-\mathcal{P}$, where $\mathcal{P}=\frac{p_{r} + 2p_{t}}{3}$ \cite{moraes2017analytical}.\\

The independent field equations corresponding to the Eqs.~(\ref{me15}), (\ref{16}), and (\ref{18}) are written as
\begin{equation}\label{19}
\frac{1}{2 r^{2}}\left[1 - \frac{b}{r}\right]\Bigg\{2r f_{QQ}Q'\left(\frac{b}{r - b}\right) + f_{Q}\left[\frac{b}{r - b}(2 + rQ') + \frac{(2r - b)(b' r - b)}{(r - b)^{2}}\right] + f\left(\frac{r^{3}}{r - b}\right)\Bigg\} 
=\rho^{eff},
\end{equation}
\begin{equation}\label{20}
-\frac{1}{2 r^{2}}\left[1 - \frac{b}{r}\right]\Bigg\{ 2r f_{QQ}Q'\left(\frac{b}{r - b}\right) + f_{Q}\left[\frac{b}{r - b}\left(2 + \frac{rb' - b}{r - b} + r \Phi'\right) - 2r\Phi'\right] + f\left(\frac{r^{3}}{r - b}\right) \Bigg\} 
=p_{r}^{eff},
\end{equation}
\begin{multline}\label{21}
-\frac{1}{4r}\left[1 - \frac{b}{r}\right]\Bigg\{-2r\Phi'f_{QQ}Q' + f_{Q}\left[2\Phi'\left(\frac{2b-r}{r - b}\right) - r\left(\Phi'\right)^{2} + \frac{rb' - b}{r(r - b)}\left(\frac{2r}{r - b} + r\phi'\right) -2r\Phi''\right]\\+ 2f\left(\frac{r^{2}}{r - b}\right)\Bigg\} 
=p_{t}^{eff},
\end{multline}
where
\begin{equation}\label{22}
\left.
\begin{split}
& \rho^{eff}= \left(8\pi + \frac{\alpha}{2} + \frac{\beta}{2}\right)\rho + \frac{\beta}{6}\left(6\rho - p_{r} - 2p_{t}\right)\\
& p_{r}^{eff}=\left(8\pi + \frac{\alpha}{2} + \frac{\beta}{2}\right)p_{r} + \frac{\beta}{6}\left(-3\rho + 4p_{r} + 2p_{t}\right)\\
& p_{t}^{eff}= \left(8\pi + \frac{\alpha}{2} + \frac{\beta}{2}\right)p_{t} + \frac{\beta}{6}\left(-3\rho + p_{r} + 5p_{t}\right)
     \end{split}
     \right \}
\end{equation}
Here $f\equiv f(Q)$,~$f_{Q}=\frac{df}{dQ}$,~and~$f_{QQ}=\frac{d^{2}f}{dQ^{2}}$.
By solving \crefrange{19}{21} and using \cref{22}, we obtain
\begin{multline}\label{23}
    \rho(r)=\frac{1}{{3 r^3 (\alpha +2 \beta +16 \pi ) (\alpha +4 \beta +16 \pi ) (r-b(r))}} \Bigg[r^2 (b'(r) f_Q \left(6 \alpha +16 \beta -\beta  r \phi '(r)+96 \pi \right)\\+\beta  r \left(2 r Q'(r) \phi '(r) f_{QQ}+f_Q \left(2 r \phi ''(r)+r \phi '(r)^2+4 \phi '(r)\right)\right))+r b(r) (b'(r) f_Q \left(-3 \alpha -10 \beta +\beta  r \phi '(r)-48 \pi \right)\\+2 r Q'(r) f_{QQ} \left(3 \alpha +8 \beta -2 \beta  r \phi '(r)+48 \pi \right)+r f_Q \left(3 (\alpha +3 \beta +16 \pi ) Q'(r)-2 \beta  \left(2 r \phi ''(r)+r \phi '(r)^2+5 \phi '(r)\right)\right))\\-b(r)^2 (2 r Q'(r) f_{QQ} \left(3 \alpha +8 \beta -\beta  r \phi '(r)+48 \pi \right)+f_Q (3 \alpha +6 \beta +3 r (\alpha +3 \beta +16 \pi ) Q'(r)-2 \beta  r^2 \phi ''(r)-\beta  r^2 \phi '(r)^2\\-6 \beta  r \phi '(r)+48 \pi ))+3 r^3 (\alpha +2 \beta +16 \pi ) (r-b(r)) f(Q)\Bigg]
\end{multline}
\begin{multline}\label{24}
    p_r(r)=\frac{1}{3 r^3 (\alpha +2 \beta +16 \pi ) (\alpha +4 \beta +16 \pi ) (r-b(r))} \Bigg[r^2 (\beta  b'(r) \left(r \phi '(r)+8\right) f_Q+r (f_Q ((6 \alpha +20 \beta +96 \pi ) \phi '(r)\\-2 \beta  r \phi ''(r)+\beta  (-r) \phi '(r)^2)-2 \beta  r Q'(r) \phi '(r) f_{QQ}))-r b(r) (f_Q (6 \alpha +b'(r) (3 \alpha +14 \beta +\beta  r \phi '(r)+48 \pi )+24 \beta\\ -3 \beta  r Q'(r)-4 \beta  r^2 \phi ''(r)-2 \beta  r^2 \phi '(r)^2+15 \alpha  r \phi '(r)+50 \beta  r \phi '(r)+240 \pi  r \phi '(r)+96 \pi )+2 r Q'(r) f_{QQ} (3 \alpha +8 \beta -2 \beta  r \phi '(r)\\+48 \pi ))+b(r)^2 (2 r Q'(r) f_{QQ} (3 \alpha +8 \beta -\beta  r \phi '(r)+48 \pi )+f_Q (9 \alpha +30 \beta -3 \beta  r Q'(r)-2 \beta  r^2 \phi ''(r)-\beta  r^2 \phi '(r)^2\\+3 r (3 \alpha +10 \beta +48 \pi ) \phi '(r)+144 \pi ))-3 r^3 (\alpha +2 \beta +16 \pi ) (r-b(r)) f(Q)\Bigg]
\end{multline}
\begin{multline}\label{25}
    p_t(r)=\frac{1}{6 r^3 (\alpha +2 \beta +16 \pi ) (\alpha +4 \beta +16 \pi ) (r-b(r))} \Bigg[r^2 (r (2 r (3 \alpha +10 \beta +48 \pi ) Q'(r) \phi '(r) f_{QQ}\\+f_Q (2 r (3 \alpha +10 \beta +48 \pi ) \phi ''(r)+r (3 \alpha +10 \beta +48 \pi ) \phi '(r)^2+2 (3 \alpha +8 \beta +48 \pi ) \phi '(r)))-b'(r) f_Q (6 \alpha +8 \beta\\ +r (3 \alpha +10 \beta +48 \pi ) \phi '(r)+96 \pi ))-r b(r) (f_Q (-6 \alpha -b'(r) \left(r (3 \alpha +10 \beta +48 \pi ) \phi '(r)-4 \beta \right)-24 \beta -6 \beta  r Q'(r)\\+12 \alpha  r^2 \phi ''(r)+6 \alpha  r^2 \phi '(r)^2+40 \beta  r^2 \phi ''(r)+20 \beta  r^2 \phi '(r)^2+192 \pi  r^2 \phi ''(r)+96 \pi  r^2 \phi '(r)^2+15 \alpha  r \phi '(r)+40 \beta  r \phi '(r)\\+240 \pi  r \phi '(r)-96 \pi )+4 r Q'(r) f_{QQ} (r (3 \alpha +10 \beta +48 \pi ) \phi '(r)-4 \beta ))+b(r)^2 (2 r Q'(r) f_{QQ} (r (3 \alpha +10 \beta +48 \pi ) \phi '(r)\\-8 \beta )+f_Q (-12 \beta -6 \beta  r Q'(r)+r^2 (3 \alpha +10 \beta +48 \pi ) \phi '(r)^2+6 \alpha  r^2 \phi ''(r)+20 \beta  r^2 \phi ''(r)+96 \pi  r^2 \phi ''(r)\\+3 r (3 \alpha +8 \beta +48 \pi ) \phi '(r)))-6 r^3 (\alpha +2 \beta +16 \pi ) (r-b(r)) f(Q)\Bigg]
\end{multline}
Here, we will explore an example that features a nonlinear function like the Starobinsky model as \cite{starobinsky2007disappearing}
\begin{equation}\label{26}
    f(Q)=Q+m Q^2,
\end{equation}
where $m$ is a constant and an unrestricted parameter. When $m=0$, the situation reduces to a linear functional case. 
%=========================================================================================================
\section{Wormhole solutions using Karmarkar condition}\label{sec4}
In this section, we want to present a well-known technique to evaluate a complete description of the wormhole configuration, called the embedding class-1 approach based on the Karmarkar condition \cite{karmarkar1948gravitational}. First, we look at the spherically symmetric spacetime metric as
\begin{equation}\label{27}
    d s^2=e^{\sigma(r)} d t^2- e^{\gamma(r)} d r^2- r^2(d \theta^2 +sin^2{\theta}d\phi^2).
\end{equation}
We next compute the Riemannian curvature components in the Schwarzschild coordinate spacetime ($t,r,\theta,\phi$) as
 \begin{align*}
 R_{1212}=\frac{-e^\sigma(2 \sigma''+{\sigma'}^2-\sigma'\gamma')}{4}, R_{2323}=-\frac{r\gamma'}{2}, R_{1414}=R_{1313}sin^2\theta,\\
 R_{3434}=\frac{r^2 sin^2 \theta e^{\gamma-1}}{e^{\gamma}}, R_{1313}=-\frac{e^\sigma(r\sigma')}{2e^\gamma}, R_{1224}=0.
 \end{align*}
Eisenhart \cite{eisenhart1997riemannian} first provided the convenient and essential requirements for embedding class-1 solutions based on the `Gauss equation’ and the `Codazzi equation’, which depend on the symmetric tensor of rank two, $b_{rs}$, as well as the Riemann curvature component $R_{pqrs}$.
\begin{itemize}
    \item The illustration of the Gauss equation is
    \begin{equation}\label{g19}
     R_{pqrs}= 2 \epsilon b_{p[r}b_{s]q}.   
    \end{equation}
    \item The expression for the Codazzi equation is
    \begin{equation}\label{g20}
     b_{p[q;r]}-\Gamma^{s}_{[qr]}b_{ps}+\Gamma^{s}_{p\;[q}b_{r]s}. 
    \end{equation}
\end{itemize}
In this case, $\epsilon=\pm1$, and the anti-symmetrization is shown by square brackets, whereas $b_{pq}$ stands for the second differential form's coefficients. By applying the aforementioned mathematical procedure to \cref{g19} and \cref{g20}
and using Pandey and Sharma condition  \cite{pandey1982insufficiency}, $R_{2323}\neq R_{1414}\neq 0$, we may write the Karmarkar Condition as
\begin{equation}\label{kar} 
R_{2323}R_{1414}=R_{1224}R_{1334} + R_{1212}R_{3434}.
\end{equation}
Now, by putting the Riemannian tensor components that we previously computed in \cref{kar}, we obtain the following.
\begin{equation}\label{28}
    [\sigma'(r)\gamma'(r)+\sigma'(r)^2-2\{\sigma''(r)+\sigma'(r)^2\}]-\frac{\sigma'(r)\gamma'(r)}{1-e^{\gamma(r)}}=0,\;\;e^{\gamma(r)}\neq 1,
\end{equation}
with primes signifying derivatives with respect to $r$.
The solution to \cref{28} yields
\begin{equation}\label{28a}
    e^{\gamma(r)}=1+D e^{\sigma(r)}\sigma'^2(r),
\end{equation}
where $D$ denotes a constant arising from integration.
Gupta et al. \cite{Gupta1975} examined the subsequent embedded Euclidean spacetime in six dimensions:
\begin{equation}\label{29}
    d s^2=-d Y^2_1-d Y^2_2-d Y^2_3+d Y^2_4+d Y^2_5\pm d Y^2_6,
\end{equation}
with their given coordinate transformation being as described below:
\begin{equation}\label{30}
\begin{aligned}
    Y_1=r\; sin\;{\theta}\; cos\; \phi,\; Y_2=r\;sin\;\theta\;sin\;\phi,\;Y_3=r\;cos\;\theta,\\
    Y_4=D\;e^{\frac{\gamma}{2}}\;cosh\;(\frac{t}{D}),\;Y_5=D\;e^{\frac{\sigma}{2}}\;sinh\;(\frac{t}{D}),\;Y_6=\;G(r),
\end{aligned}
\end{equation}\label{31}
given $G'^2(r)=\mp[(-e^\gamma-1)+\frac{D^2e^\sigma\sigma'^2}{4}]$, assuming that $D$ is positive. With the help of the metric \cref{27} and the transformation (\ref{30}), we construct an embedded class-1 spacetime that satisfies the Karmarkar criterion.
\par
The following relationships between gravitational features can be found through the comparison of \cref{me15} with \cref{27}:
\begin{equation}\label{32}
    \sigma(r)=\Phi(r),\;\; \gamma(r)=Log\bigg[\frac{r}{r-b(r)}\bigg].
\end{equation}
We now need to take into account a particular type of redshift function in order to produce the embedded class one solution, as \cite{PhysRevD.57.829, Shamir:2017byl}
\begin{equation}\label{33}
\sigma(r)=\Phi(r)=-\frac{2\mu}{r},
\end{equation}
where $\mu$ could be any arbitrary constant.
It should be noticed that the no-horizon criterion stated in condition (\ref{cond1}) of the previous section is entirely satisfied by this specific form of the redshift function $\sigma(r)$. Now, we may solve the consequent equation by replacing \cref{32} and \cref{33} in \cref{28a}. This gives us
\begin{equation}\label{34}
b(r)=r-\frac{r^{5}}{r^{4} + 4 \mu^{2} D e^{-2\mu/r}}
\end{equation}
This shape function $b(r)$ satisfies every traversability requirement covered in the preceding section. Therefore, condition (\ref{cond2}) $b(r_t)=r_t$, where $r_t$ is the wormhole throat radius, is likewise met.
Although the application of the throat condition results in only a trivial solution. This problem is resolved by adding a free parameter $P$ to \cref{34}, which is written as follows:
\begin{equation}\label{35}
b(r)=r-\frac{r^{5}}{r^{4} + 4 \mu^{2} D e^{-2\mu/r}}+P.
\end{equation}
Next, in order to remove the integration constant $D$, we once again apply the throat condition, $b(r_t)=r_t$, and following a successful removal, \cref{35} provides
\begin{equation}\label{36}
b(r)=P -\frac{P  r^5}{P  r^4+r^4_t (r_t-P) e^{\frac{2 \mu  (r-r_t)}{r r_t}}}+r,
\end{equation}
with $0<P<r_t$.
\par
From now on, the function $b(r)$ will be treated as the shape function of the wormhole wherever it is needed.
\par
\begin{figure}[!h]
  \centering
  \includegraphics[width=0.4\textwidth]{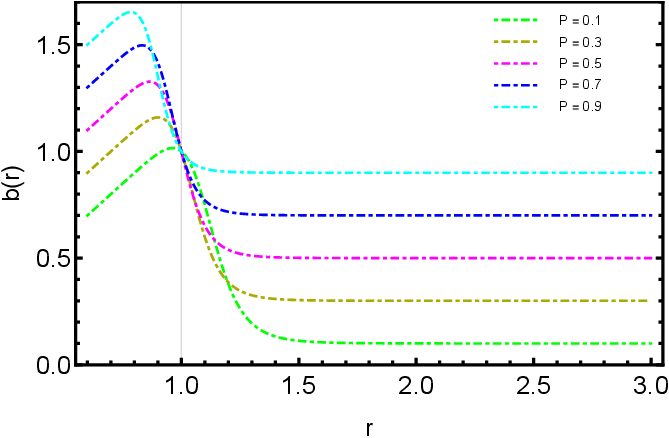}
  \hspace{5mm}
  \vspace{5mm}
  \includegraphics[width=0.4\textwidth]{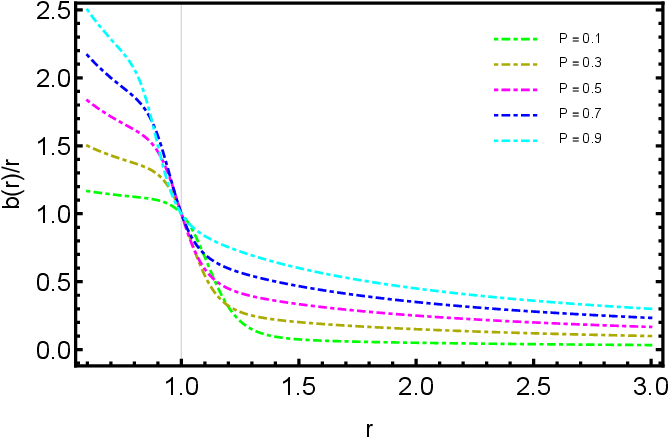}
  \vspace{5mm}
  \includegraphics[width=0.4\textwidth]{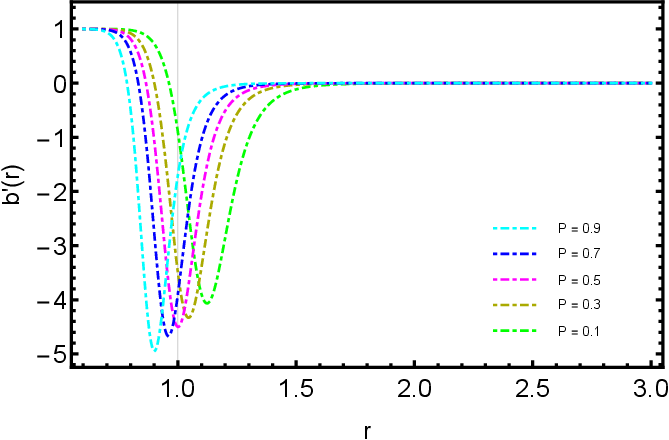}
  \hspace{3mm}
  \includegraphics[width=0.4\textwidth]{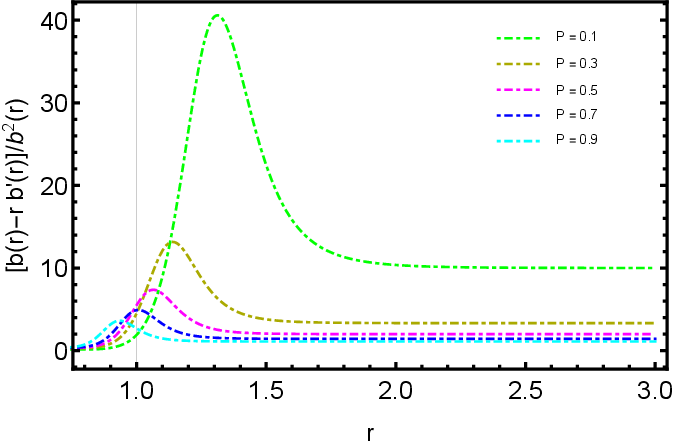}
  \includegraphics[width=0.4\textwidth]{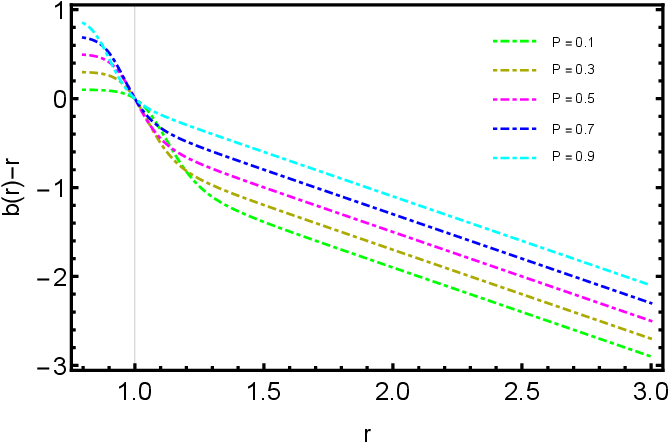}
  \hspace{5mm}
  \includegraphics[width=0.4\textwidth]{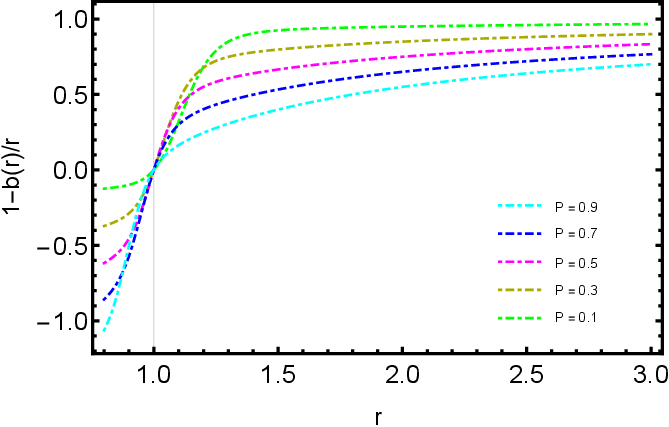}
  \caption{Plots of the shape function b(r) and related expressions for various values of the parameter $P$, with throat radius $r_t=1$ kpc and  $\mu=-8$, highlighting the wormhole's geometric features. }\label{fig1}
\end{figure}

Prior to that, we aim to examine the characteristics of this function in more detail. To this end, we utilize graphical methods, and \cref{fig1} presents different curves corresponding to the shape function $b(r)$ for certain selected values of the constant $P$. Upon careful examination of each graphical profile, it is evident that the shape function obtained in this study adheres to the fundamental geometric requirements for a wormhole, including the presence of a throat, the flaring-out behavior, and asymptotic flatness, as outlined in conditions (\ref{cond2}) to (\ref{cond5}).

%%%%%%%%%%%%%%%%%%%%%%%%%%%%%%%%%%%%%%%%%%%%%%%%%%%%%%%%%%%%%%%%%%%%%%%%
%==========================================================================================
\section{Embedding diagram}\label{sec5}

To aid in visualizing the wormhole configuration supplied by \cref{me15}, the shape function, as defined in \cref{36}, is represented by embedded diagrams of two and three dimensions. In three-dimensional cylindrical coordinates $(r, \phi, z)$, where $z$ represents the necessary embedding surface, we want to examine the structure of our wormhole scenario. In light of this, we take into account the two-dimensional hypersurface $\Lambda$, which is provided by $\Lambda=\{(t,r,\theta,\phi\}:\theta=\frac{\pi}{2},t=constant\}$. Consequently, the line element (\ref{me15}) can be expressed in the modified form as
\begin{equation}\label{37}
    d s^2_\Lambda=- \frac{dr^2}{1-\frac{b(r)}{r}}-r^2 d\phi^2.
\end{equation}
Then, it is possible to embed the preceding equation \cref{37} in cylindrical coordinates $(r, \phi, z)$ as
\begin{equation}\label{38}
    d s^2_\Lambda=-dr^2-dz^2-r^2 d\phi^2.
\end{equation}
Hence, within three-dimensional Euclidean space, this metric may be appropriately adjusted as
\begin{equation}\label{39}
d s^2_\Lambda=-\bigg[1+\bigg(\frac{dz}{dr}\bigg)^2\bigg]dr^2-r^2d\phi^2.  \end{equation}
With the help of \cref{37} and \cref{39}, we get
\begin{equation}
    \frac{dz}{dr}={\pm}{\sqrt{\Bigg(1-\frac{b(r)}{r}\Bigg)^{-1}-1}}.
\end{equation}
The integration of which over the radial coordinate $r$ from $r_t$ to $r$ yields
\begin{equation}\label{44}
    \begin{aligned}
    z(r)&~=~{\pm}\int^r_{r_t}{\sqrt{\Bigg(1-\frac{b(r)}{r}\Bigg)^{-1}-1}}
        &~=~{\pm}\int^r_{r_t}{\sqrt{\frac{r}{P  \left(\frac{r^5}{P  r^4+r^4_t (r_t-P ) e^{2 \mu  \left(\frac{1}{r_t}-\frac{1}{r}\right)}}-1\right)}-1}}\;\; dr.
    \end{aligned}
\end{equation}
This represents the necessary expression associated with the embedding surface $z(r)$.

\begin{figure}[!h]
  \centering
  \includegraphics[width=0.43\textwidth]{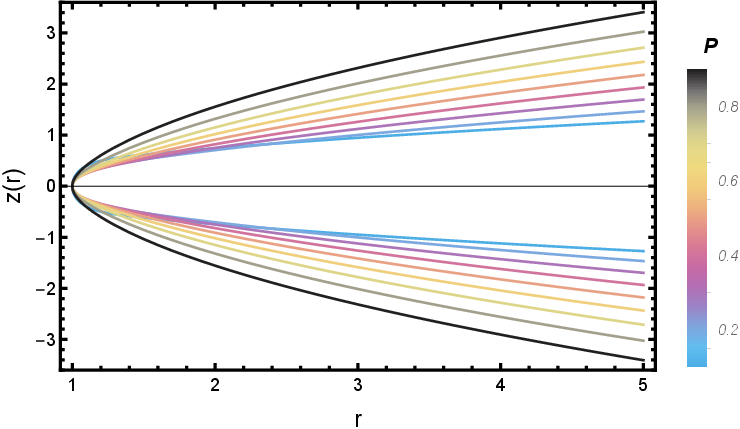}
  \hspace{5mm}
  \includegraphics[width=0.45\textwidth]{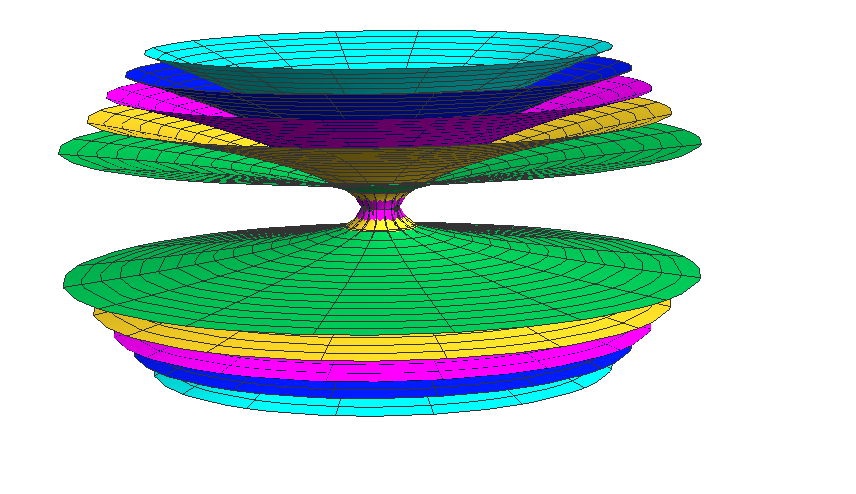}
  \caption{Embedding two-dimensional diagram (left panel) for various values of the parameter $P\in[0.1,0.9]~\text{with step size 0.1}$ and three-dimensional diagram (right panel) corresponding to $5$ values of $P=0.1\rightarrow$ \textcolor{green}{$\spadesuit$}, $P=0.3\rightarrow$ \textcolor{yellow}{$\spadesuit$}, $P=0.5\rightarrow$ \textcolor{violet}{$\spadesuit$}, $P=0.7\rightarrow$ \textcolor{blue}{$\spadesuit$}, $P=0.9\rightarrow$ \textcolor{cyan}{$\spadesuit$} with throat radius $r_t=1$ kpc and  $\mu=-8$.}\label{fig3.1}
\end{figure}
\begin{figure}[!h]
  \centering
  \includegraphics[width=0.50\textwidth]{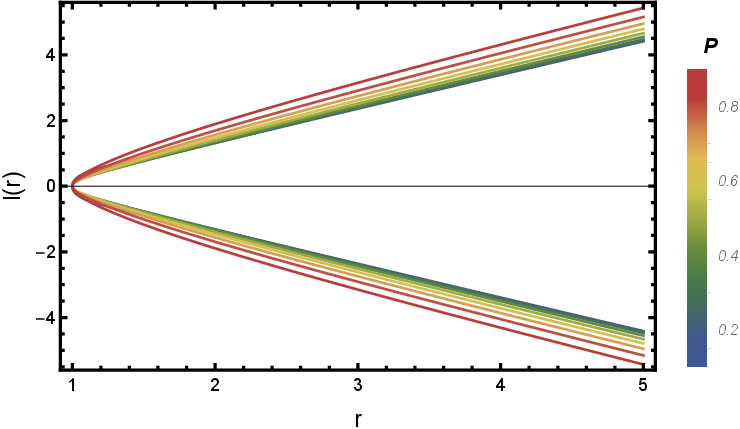}
  \caption{Graphical representation of $l(r)$ as a function of $r$ for various values of the parameter $P\in[0.1,0.9]~\text{with step size 0.1}$ with throat radius $r_t=1$ kpc and  $\mu=-8$.}\label{fig3.2}
\end{figure}
A method introduced by Kruskal \cite{Kruskal:1959vx} and Szekeres \cite{Szekeres:1960gm, Szekeres2002} in 1960 was regarded as the most extensive extension of the Schwarzschild solution, developed in response to the discovery of the Einstein-Rosen bridge. They further developed the Kruskal coordinate system, which is essential for analyzing and interpreting the embedding diagram. In the current analysis, we employ the approach initially introduced by Marolf \cite{Marolf:1998dz} and further developed by Collas and Klein \cite{Collas:2011py} to construct the wormhole embedding diagram. In \cref{fig3.1}, we illustrate both 2D and 3D representations of the embedded wormhole surface given by \cref{44}, corresponding to different selected values of the constant $P$.
%====================================================================================================
\section{proper radial distance}\label{sec6}
According to references \cite{Lobo:2017cay, Ghosh:2021dgm}, using the proper radial coordinate, the metric in \cref{me15} may be rewritten as follows:
\begin{equation}
    ds^2=e^{\Phi(l)}dt^2-dl^2-r(l)^2(d\theta^2+sin^2\theta d\phi^2,
\end{equation}
where $l$ is determined by
\begin{equation}
    \begin{aligned}
    l(r)&~=~{\pm} \int_{r_{t}}^r (1-\frac{b(r)}{r})^{-\frac{1}{2}} dr
        &~=~{\pm} \int_{r_{t}}^r \frac{1}{\sqrt{\frac{P  \left(\frac{r^5}{P  r^4+r^4_t (r_t-P ) e^{2 \mu  \left(\frac{1}{r_t}-\frac{1}{r}\right)}}-1\right)}{r}}} \;\; dr.
    \end{aligned}
\end{equation}
The function $l(r)$ represents the proper radial distance within the wormhole geometry, while the symbol '$\pm$' signifies the presence of its two mouths. For a traversable wormhole to be accurately described, $l(r)$ must be well-behaved and finite throughout the entire spacetime. Moreover, the proper radial distance $l(r)$ should satisfy the condition that it is at least as large as the absolute value of the coordinate separation, which is expressed as $|l(r)|\geq|r-r_{t}|$ \cite{Nandi:1998ie, Das:2024qmi}. This fundamental condition is analyzed in \cref{fig3.2}, where the behavior of $l(r)$ is plotted against the increasing radial coordinate $r\geq{r^+_t}$. \\

It is essential that $l(r)$ remains greater than or equal to the corresponding radial coordinate to ensure a physically valid embedding of the wormhole. \cref{table1} provides computed values of $z(r)$ and $l(r)$ for several points in the domain $r>r_t$, focusing on the first quadrant.  From these values, it is evident that $l(r)$ consistently remains slightly larger than $z(r)$ at each point, thus satisfying the necessary geometric condition. This confirms that the embedding structure of the proposed wormhole is smooth and well-behaved throughout the considered spatial region.\\\\ 

\begin{xltabular}{\linewidth}{>{\scriptsize}p{0.03\textwidth} >{\scriptsize}p{0.07\textwidth} >{\scriptsize}p{0.07\textwidth} >{\scriptsize}p{0.07\textwidth} >{\scriptsize}p{0.07\textwidth} >{\scriptsize}p{0.07\textwidth} >{\scriptsize}p{0.001\textwidth} >{\scriptsize}p{0.07\textwidth}>{\scriptsize}p{0.07\textwidth}>{\scriptsize}p{0.07\textwidth}>{\scriptsize}p{0.07\textwidth}>{\scriptsize}p{0.07\textwidth}} 

\caption{\centering\textit{Tabulated approximations of the functions $z(r)$ and $l(r)$ computed for the case where the wormhole throat radius is $r_{t}=1$ kpc. and the parameter $\mu=-8$.}}
\label{table1}\\

    \toprule\toprule
   $\textbf{\large{r}}$ & $\textbf{\large{z(r)}}$ &&&&&& $\textbf{\large{l(r)}}$ \\ \addlinespace
   \cline{2-6}\cline{8-12} & $P = 0.1$ & $P = 0.3$ & $P = 0.5$ & $P = 0.7$ & $P = 0.9$  && $P = 0.1$ & $P = 0.3$ & $P = 0.5$ & $P = 0.7$ & $P = 0.9$
    \\\addlinespace
    
    \toprule\toprule 
    {$1.01$} & {$0.14304$} & {$0.09315$} & {$0.08465$} & {$0.09029$} & {$0.12244$} && {$0.14350$} & {$0.09386$} & {$0.08543$} & {$0.09102$} & {$0.122985$} \\
    {$1.5$} & {$0.61737$} & {$0.52125$} & {$0.59289$} & {$0.73940$} & {$1.03545$} && {$0.88287$} & {$0.76828$} & {$0.80487$} & {$0.91405$} & {$1.16941$} \\
    {$2$} & {$0.74292$} & {$0.74993$} & {$0.91089$} & {$1.1507$} & {$1.55575$} && {$1.39844$} & {$1.3182$} & {$1.39764$} & {$1.56186$} & {$1.8917$} \\
    {$2.5$} & {$0.85102$} & {$0.94646$} & {$1.17885$} & {$1.48781$} & {$1.96578$} && {$1.91001$} & {$1.85548$} & {$1.965$} & {$2.16503$} & {$2.53855$} \\
    {$3$} & {$0.94827$} & {$1.12166$} & {$1.41492$} & {$1.78054$} & {$2.31533$} && {$2.41938$} & {$2.38531$} & {$2.51797$} & {$2.74449$} & {$3.14872$} \\
    {$3.5$} & {$1.03743$} & {$1.28125$} & {$1.62834$} & {$2.04282$} & {$2.62519$} && {$2.92728$} & {$2.91017$} & {$3.06164$} & {$3.30914$} & {$3.73701$} \\
    {$4$} & {$1.12024$} & {$1.42879$} & {$1.8246$} & {$2.28256$} & {$2.90644$} && {$3.43409$} & {$3.43149$} & {$3.5988$} & {$3.86367$} & {$4.31072$} \\
    {$4.5$} & {$1.19789$} & {$1.56665$} & {$2.00728$} & {$2.50472$} & {$3.16578$} && {$3.94008$} & {$3.95016$} & {$4.13113$} & {$4.41082$} & {$4.87399$} \\
    {$5$} & {$1.27124$} & {$1.69653$} & {$2.17885$} & {$2.71269$} & {$3.40765$} && {$4.44544$} & {$4.46675$} & {$4.65976$} & {$4.95236$} & {$5.42944$} \\
\bottomrule
\end{xltabular}
%========================================================================================
\section{Energy conditions}\label{sec7}

Several inequalities related to the energy–momentum tensor that control the energy density, the radial pressure, and the tangential pressure are essential for the existence and stability of the wormhole. These inequalities, commonly known as energy conditions, play a vital role in identifying wormhole solutions. Energy circumstances are essential and powerful principles to predict diverse behaviors related to black holes, wormholes, and the geodesic paths within the Universe, as well as additional physical situations. The Raychaudhuri equations provide the basis for these requirements \cite{Raychaudhuri:1953yv, Nojiri:2006ri, Ehlers:2006aa}. The following is the formulation of the Raychaudhuri equations:
\begin{equation}
    \frac{d \Psi}{d \tau}=-\frac{\Psi^2}{3}-\sigma_{pq}\sigma^{pq} +\Theta_{pq} \Theta^{pq}-R_{pq}u^p u^q,
\end{equation}
\begin{equation}
    \frac{d \Psi}{d \tau}=-\frac{\Psi^2}{2}-\sigma_{pq}\sigma^{pq} +\Theta_{pq} \Theta^{pq}-R_{pq}v^p v^q,
\end{equation}
where $\Psi$, $\Theta$, and $\sigma$ denote the expansion, rotation, and shear of the congruence associated with the vector field $u^p$.
\par
As our present investigation solely addresses the anisotropic distribution of matter, four distinct and widely recognized energy conditions have been examined and can be stated as follows:
\begin{enumerate}
    \item Null energy condition (NEC): $\rho+p_m\geq0$ for all $m$, specifically conditions $\rho+p_r\geq0$ and $\rho+p_t\geq0$ must be satisfied.
    \item Weak energy condition (WEC): $\rho\geq0$, $\rho+p_m\geq0$ for all $m$.
    \item Strong energy condition (SEC): $\rho+p_m\geq0$ for all $m$ and $\rho+\sum_m p_m\geq0$ for all $m$,  specifically $\rho+p_r+2 p_t\geq0$.
    \item Dominant energy condition (DEC): $\rho\geq0$ and $\rho-|p_m|\geq0$ for all $m$.
\end{enumerate}
%%%%%%%%%%%%%%%%%%%%%%%%%%%%%%%%%%%%%%%%%%%%%%%%%%%%%%%%%%%%%%%%%%%%%%%%
\begin{figure}[!h]
  \centering
  \includegraphics[width=0.45\textwidth]{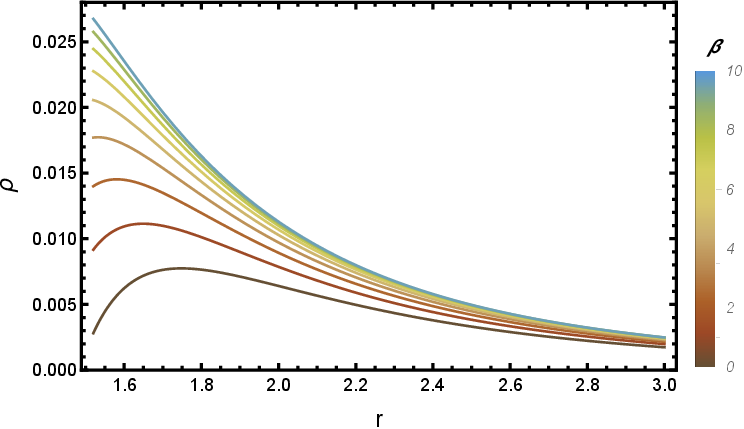}
  \hspace{5mm}
  \vspace{5mm}
  \includegraphics[width=0.45\textwidth]{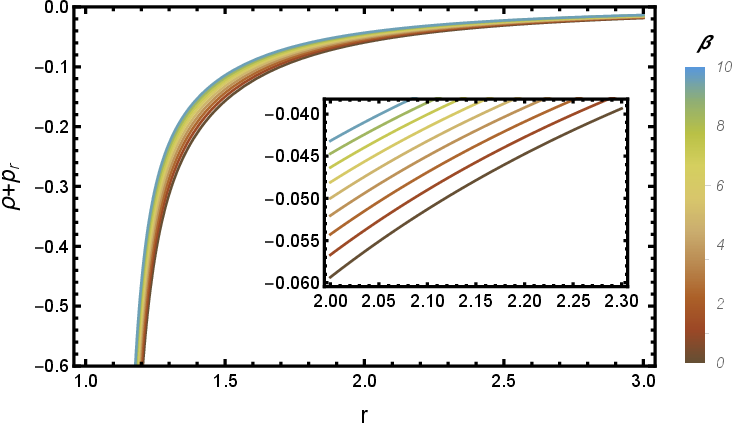}
  \vspace{5mm}
  \includegraphics[width=0.45\textwidth]{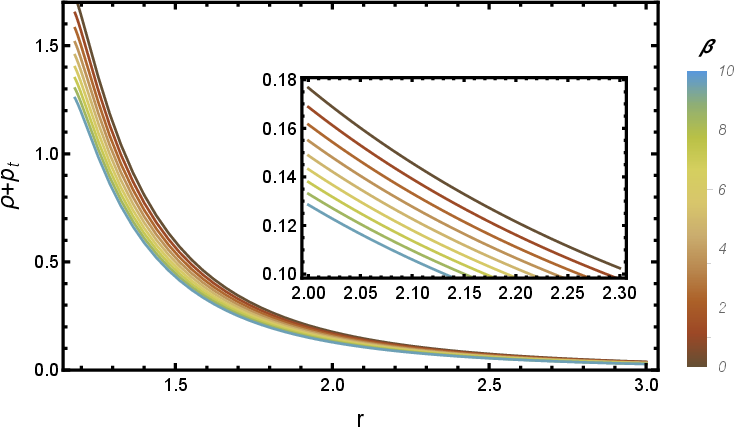}
  \hspace{3mm}
  \includegraphics[width=0.45\textwidth]{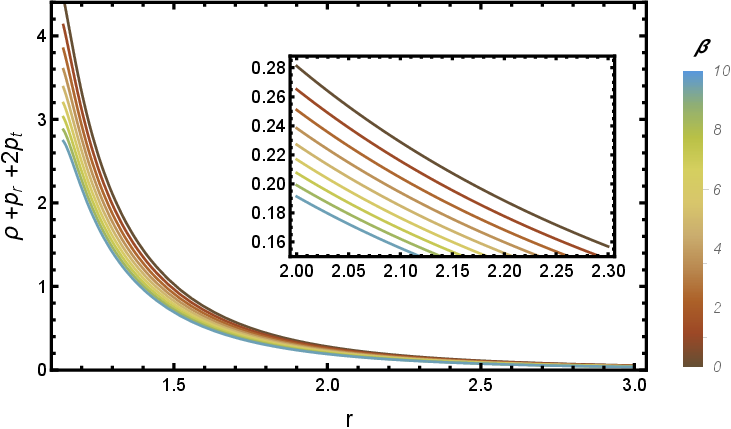}
  \includegraphics[width=0.45\textwidth]{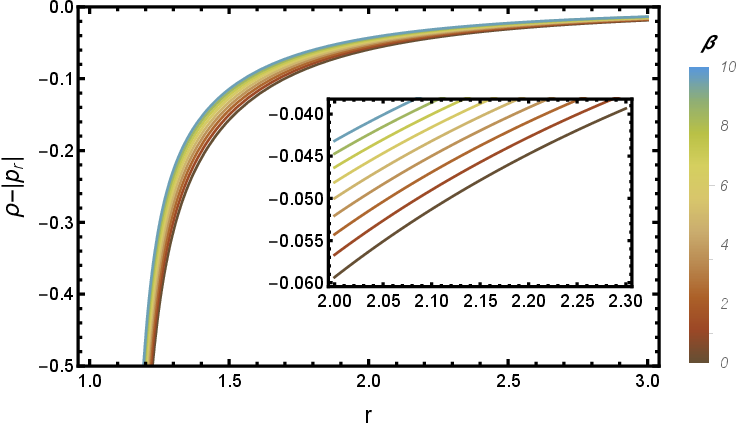}
  \hspace{5mm}
  \includegraphics[width=0.45\textwidth]{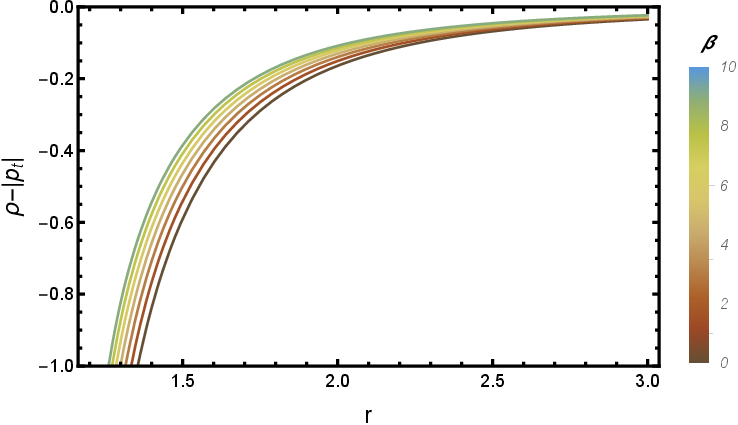}
  \caption{Energy condition expressions $\rho$ (above left), $\rho+p_r$ (above right), $\rho+p_t$ (middle left), $\rho+p_r+2 p_t$ (middle right), $\rho-|p_r|$ (below left), $\rho-|p_t|$ (below right) plotted against the radial coordinate $r$ for various values of parameter $\beta\in[0,10]~\text{with step size 1.2}$ with throat radius $r_t = 1$ kpc, $\mu = -8$, $\alpha=1$, $m=0.1$, $P=0.5$.}\label{fig4.1}
\end{figure}
%%%%%%%%%%%%%%%%%%%%%%%%%%%%%%%%%%%%%%%%%%%%%%%%%%%%%%%%%%%%%%%%%%%%%%%%%
\begin{figure}[!h]
  \centering
  \includegraphics[width=0.45\textwidth]{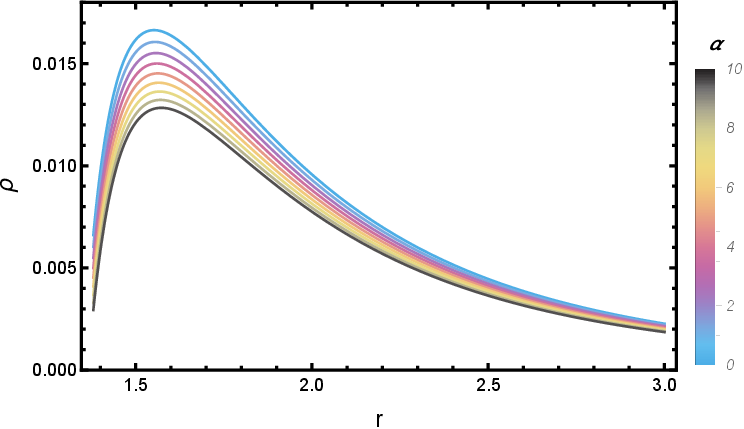}
  \hspace{5mm}
  \vspace{5mm}
  \includegraphics[width=0.45\textwidth]{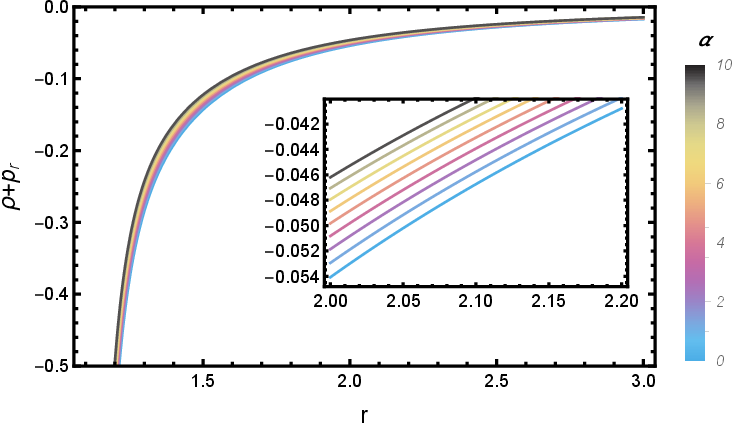}
  \vspace{5mm}
  \includegraphics[width=0.45\textwidth]{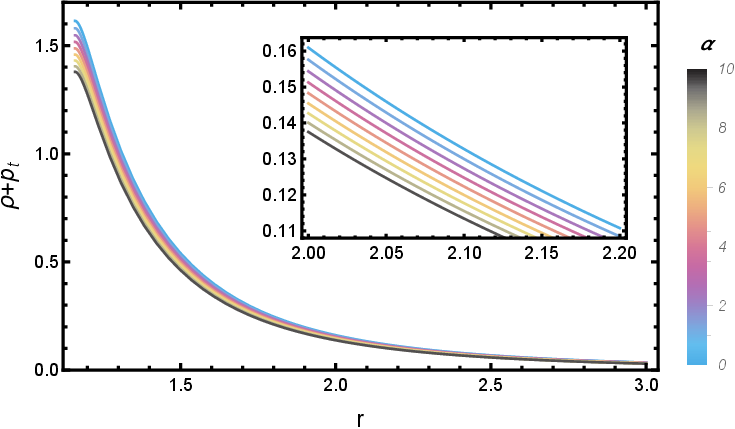}
  \hspace{3mm}
  \includegraphics[width=0.45\textwidth]{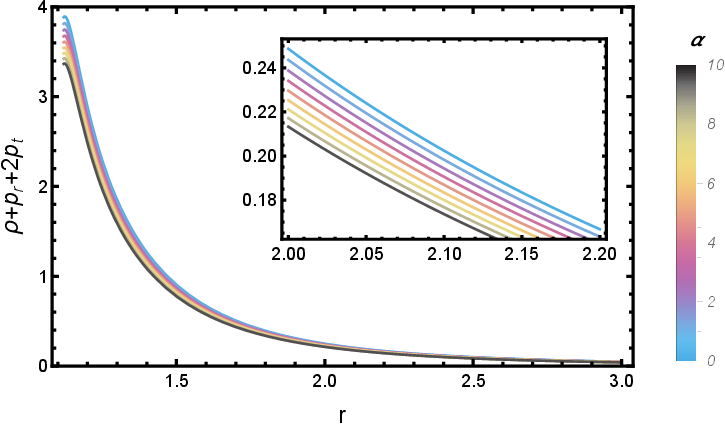}
  \includegraphics[width=0.45\textwidth]{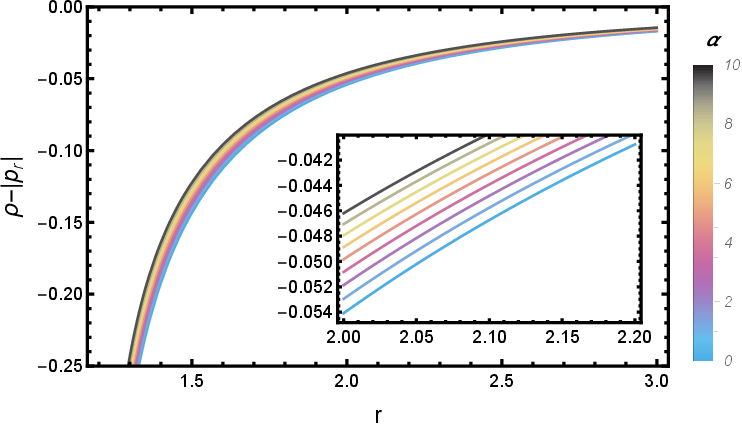}
  \hspace{5mm}
  \includegraphics[width=0.45\textwidth]{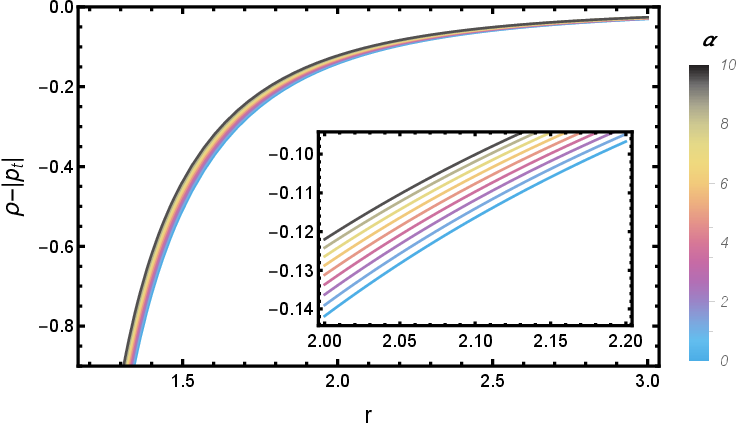}
  \caption{Energy condition expressions $\rho$ (above left), $\rho+p_r$ (above right), $\rho+p_t$ (middle left), $\rho+p_r+2 p_t$ (middle right), $\rho-|p_r|$ (below left), $\rho-|p_t|$ (below right) plotted against the radial coordinate $r$ for various values of parameter $\alpha\in[0,10]~\text{with step size 1.2}$ with throat radius $r_t = 1$ kpc, $\mu = -8$, $\beta=3$, $m=0.1$, $P=0.5$.}\label{fig4.2}
\end{figure}
%%%%%%%%%%%%%%%%%%%%%%%%%%%%%%%%%%%%%%%%%%%%%%%%%%%%%%%%%%%%%%%%%%%%%%%%%
\begin{figure}[!h]
  \centering
  \includegraphics[width=0.45\textwidth]{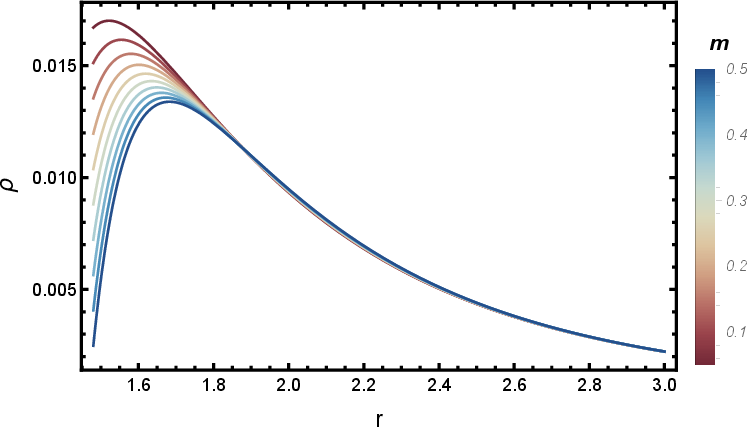}
  \hspace{5mm}
  \vspace{5mm}
  \includegraphics[width=0.45\textwidth]{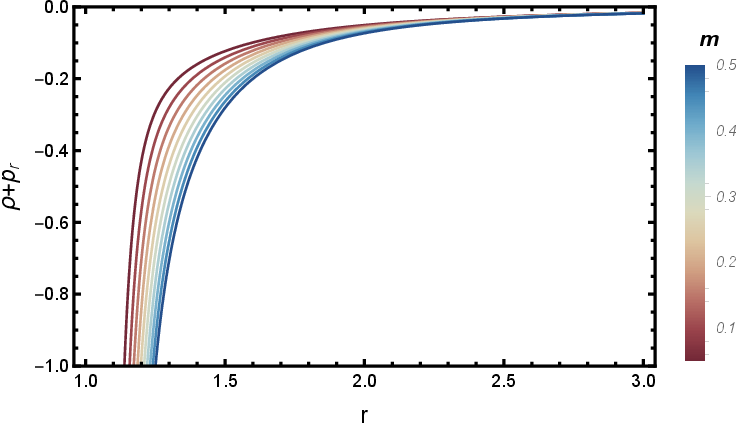}
  \vspace{5mm}
  \includegraphics[width=0.45\textwidth]{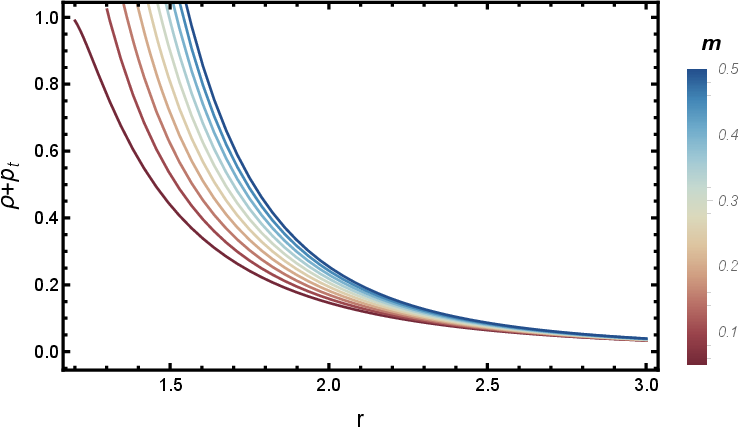}
  \hspace{3mm}
  \includegraphics[width=0.45\textwidth]{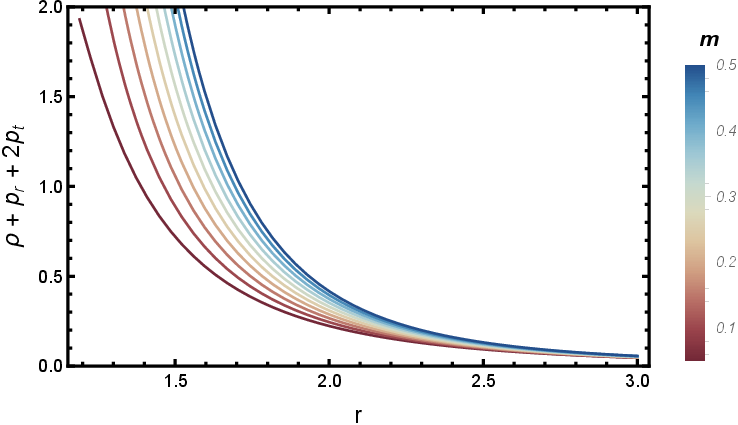}
  \includegraphics[width=0.45\textwidth]{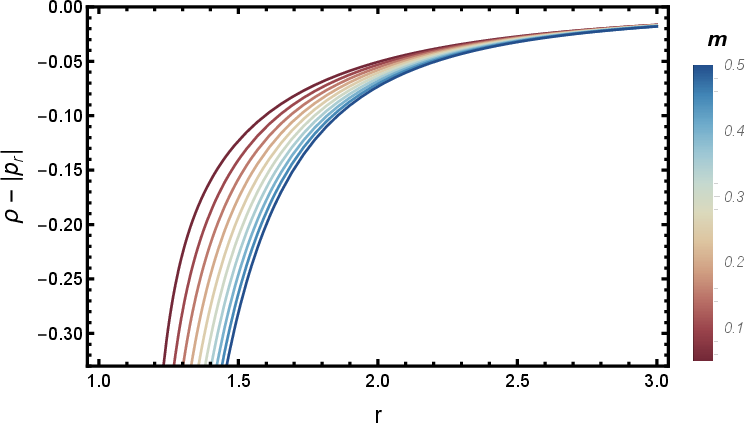}
  \hspace{5mm}
  \includegraphics[width=0.45\textwidth]{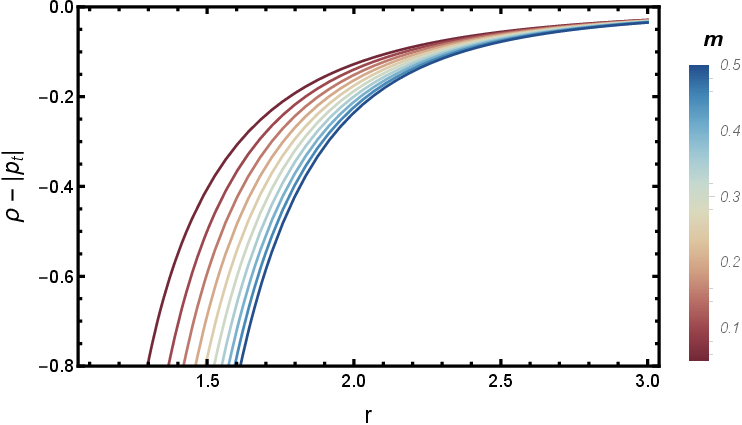}
  \caption{Energy condition expressions $\rho$ (above left), $\rho+p_r$ (above right), $\rho+p_t$ (middle left), $\rho+p_r+2 p_t$ (middle right), $\rho-|p_r|$ (below left), $\rho-|p_t|$ (below right) plotted against the radial coordinate $r$ for various values of parameter $m\in[0.05,0.5]~\text{with step size 0.05}$ with throat radius $r_t = 1$ kpc, $\mu = -8$, $\alpha=1$, $\beta=3$, $P=0.5$.}\label{fig4.3}
\end{figure}
%%%%%%%%%%%%%%%%%%%%%%%%%%%%%%%%%%%%%%%%%%%%%%%%%%%%%%%%%%%%%%%%%%%%%%%%
\begin{figure}[!h]
  \centering
  \includegraphics[width=0.45\textwidth]{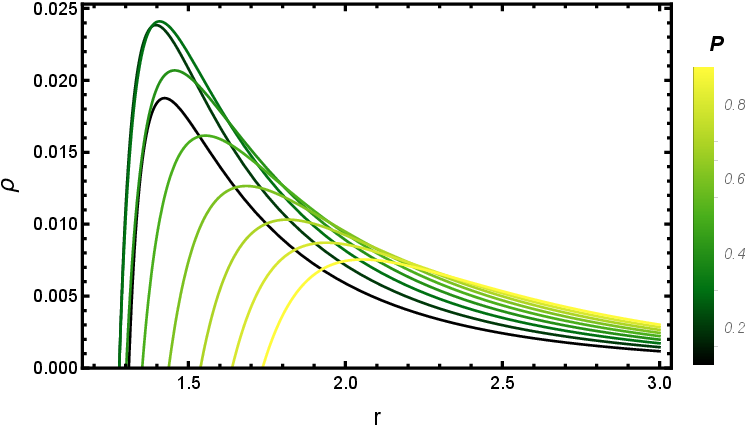}
  \hspace{5mm}
  \vspace{5mm}
  \includegraphics[width=0.45\textwidth]{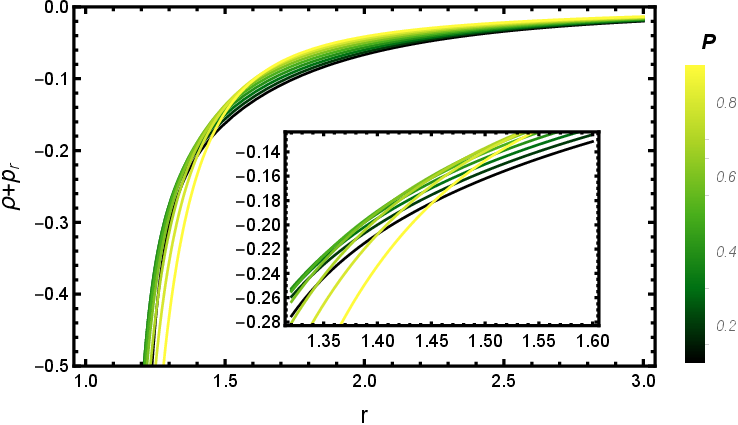}
  \vspace{5mm}
  \includegraphics[width=0.45\textwidth]{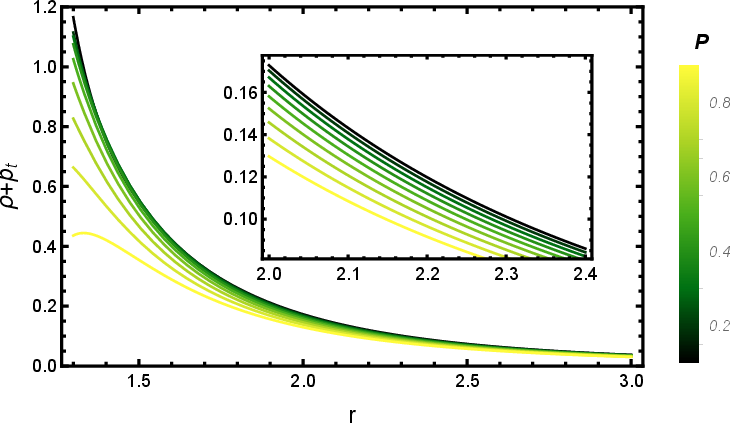}
  \hspace{3mm}
  \includegraphics[width=0.45\textwidth]{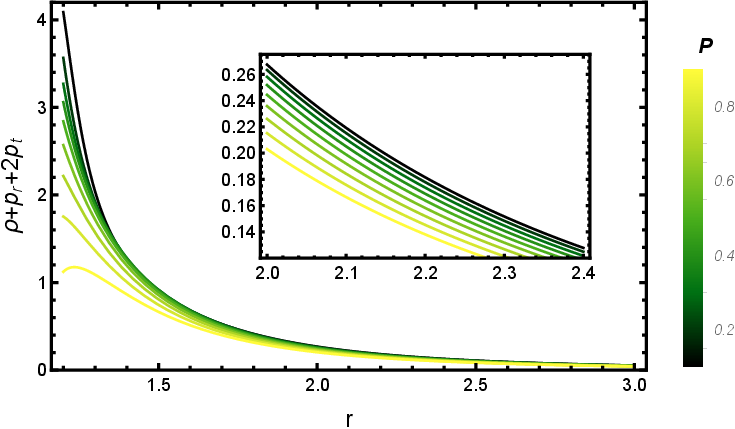}
  \includegraphics[width=0.45\textwidth]{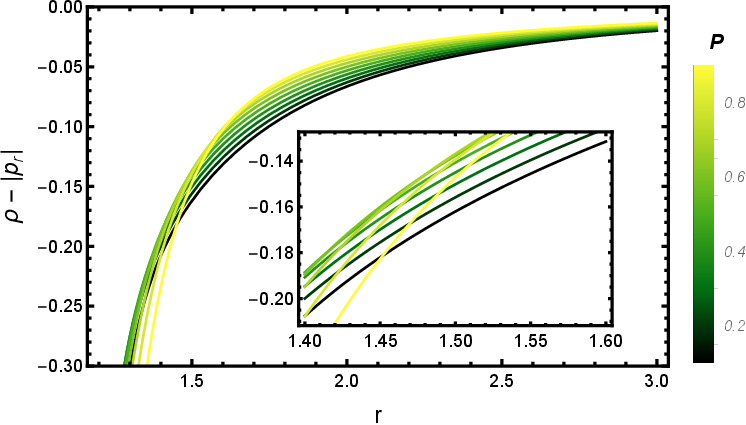}
  \hspace{5mm}
  \includegraphics[width=0.45\textwidth]{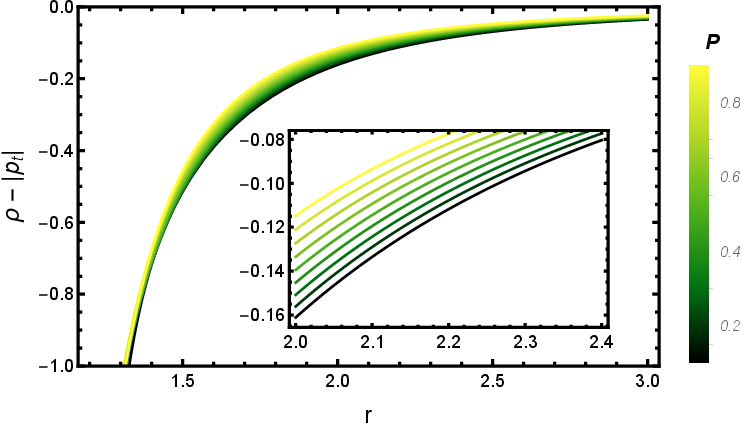}
  \caption{Energy condition expressions $\rho$ (above left), $\rho+p_r$ (above right), $\rho+p_t$ (middle left), $\rho+p_r+2 p_t$ (middle right), $\rho-|p_r|$ (below left), $\rho-|p_t|$ (below right) plotted against the radial coordinate $r$ for various values of parameter $P\in[0.1,0.9]~\text{with step size 0.1}$ with throat radius $r_t = 1$ kpc, $\mu = -8$, $\alpha=1$, $\beta=3$, $m=0.1$.}\label{fig4.4}
\end{figure}
%%%%%%%%%%%%%%%%%%%%%%%%%%%%%%%%%%%%%%%%%%%%%%%%%%%%%%%%%%%%%%%%%%%%%%%%%
\par
In the framework of GR, energy conditions typically imply that the energy density within a given region of spacetime is non-negative. These conditions are not physical laws but are instead introduced as mathematical boundary assumptions. The SEC, in physical terms, suggests that gravity is fundamentally attractive in nature and may characterize regions with strong gravitational influence. It is also important to note that if the NEC is violated, the WEC and SEC are typically violated as well. However, the converse does not necessarily hold true. In GR, a defining property of a traversable wormhole supported by exotic matter is its violation of the NEC. Moreover, a violation of the SEC is necessary to describe the geometric structure of a wormhole. Hence, when the NEC is violated, it generally results in a breakdown of the remaining fundamental energy conditions, too.
\par
Energy conditions are essential not only for characterizing wormhole configurations but also for understanding broader gravitational phenomena, including singularity theorems \cite{Hawking:1973uf}, black hole thermodynamics, and early-universe bounce models \cite{Novello:1979ik, Novello:1992tb, DeLorenci:2002mi, Singh:2015bca, Peter:2001fy}. The frequent violation of the WEC in such scenarios suggests a deeper link between wormholes and bouncing cosmologies. Additionally, the physical or exotic nature of matter inside compact astrophysical objects is often judged through these energy conditions.
\par

For the purpose of studying the nature of the energy conditions in our wormhole model, we perform a graphical analysis of six key expressions associated with $\rho$, $\rho+p_r$, $\rho+p_t$, $\rho+p_r+2p_t$, $\rho-|p_r|$, $\rho-|p_t|$. These expressions are plotted against the radial coordinate $r$ for varying values of the model parameters.  In particular, we study the effect of each parameter, namely $\beta$, $\alpha$, $m$, and $P$, while keeping the other free parameters fixed at $r_t=1$ and $\mu=-8$, on the satisfaction or violation of the energy conditions. In each case, we vary one parameter at a time to isolate its specific effect on the energy conditions. This approach enables us to assess the extent to which each parameter influences the satisfaction or violation of the Null, Weak, Strong, and Dominant Energy Conditions, which are critical in determining the physical admissibility of the wormhole solutions within the $\mathcal{F}(Q,\mathcal{L}_{m},\mathcal{T})$ gravity framework.
\par
\begin{itemize}
    \item \textbf{Case I: Variation with $\beta$ (see \cref{fig4.1})}: In this case, we analyze the impact of varying the parameter $\beta$ on the energy conditions, keeping $\alpha$, $m$, and $P$ fixed, along with $r_t=1$ and $\mu=-8$. As shown in \cref{fig4.1}, the energy density $\rho$ remains positive for $r\geq 1.4$, indicating a physically reasonable matter content in that region. However, the radial component of the NEC, $\rho+p_r$, is found to be negative throughout the domain, signaling a clear violation and, consequently, the WEC. In contrast, $\rho+p_t$ and $\rho+p_r+2p_t$ maintain positive values for larger $r$, suggesting partial satisfaction of the SEC. Both expressions $\rho-|p_r|$ and $\rho-|p_t|$ remain negative, confirming violation of the DEC. Hence, the energy conditions—NEC, WEC, SEC, and DEC—are not fully satisfied in the vicinity of the wormhole throat for the considered embedded wormhole solution in $\mathcal{F}(Q,\mathcal{L}_{m},\mathcal{T})$ gravity, indicating the unavoidable presence of exotic matter in this region.
    \item \textbf{Case II: Variation with $\alpha$ (see \cref{fig4.2})}: \cref{fig4.2} illustrates the behavior of the expressions of the energy conditions as the parameter $\alpha$ is varied, while $\beta$, $m$, and $P$ are fixed constant. Similarly to the previous case, $\rho$ stays positive beyond a certain radial distance, ensuring a non-negative energy density. The NEC in the radial direction and the WEC continue to be violated, as $\rho+p_r<0$ throughout the plotted range. Meanwhile, $\rho+p_t$ and $\rho+p_r+2p_t$ show positive profiles, implying that the SEC can be conditionally satisfied. The negative values of $\rho-|p_r|$ and $\rho-|p_t|$ again reflect a persistent violation of the DEC. Hence, the analysis reveals that none of the energy conditions, NEC, WEC, SEC, or DEC, are fully upheld for wormhole configurations when the parameter $\alpha$ is varied, especially near the throat region.
    \item \textbf{Case III: Variation with $m$ (see \cref{fig4.3})}: The behavior of the energy conditions under variation of the parameter $m$ is presented in \cref{fig4.3}. The energy density $\rho$ remains positive for $r>1.36$, while $\rho+p_r$ stays negative, confirming the violation of both NEC and WEC. Although $\rho+p_t$ remains positive in the domain, the overall WEC is not satisfied due to the negativity of the radial term. The SEC term $\rho+p_r+2p_t$ displays positive values for moderate and large $r$, suggesting a limited fulfillment of the SEC.  The DEC is again violated for all values of $m$. Consequently, the wormhole solutions evaluated in $\mathcal{F}(Q,\mathcal{L}_{m},\mathcal{T})$ gravity do not adequately fulfill the standard energy conditions, NEC, WEC, SEC, and DEC, in the neighborhood of the throat region for different values of the parameter $m$.
    \item \textbf{Case IV: Variation with $P$ (see \cref{fig4.4})}: In \cref{fig4.4}, the expressions of the energy conditions are evaluated for varying $P$, keeping the other parameters fixed. The energy density $\rho$ becomes positive for $r>1.38$, while $\rho+p_r$ remains negative, implying NEC and WEC violations. Although $\rho+p_t$ and $\rho+p_r+2p_t$ are positive in a large region, they do not compensate for the failed radial component in the WEC. The DEC continues to be violated, as both $\rho-|p_r|$ and $\rho-|p_t|$ are negative across the domain. Therefore, within the framework of $\mathcal{F}(Q,\mathcal{L}_{m},\mathcal{T})$ gravity, the wormhole solutions examined fail to satisfy the conventional energy conditions, NEC, WEC, SEC, and DEC, near the throat region when the parameter $P$ is varied.
\end{itemize}
Hence, the violation of all standard energy conditions, NEC, WEC, SEC, and DEC, for all considered values of the parameters indicates that the existence of exotic matter near the wormhole throat is inevitable within the framework of $\mathcal{F}(Q,\mathcal{L}_{m},\mathcal{T})$ gravity.

%===============================================================================================
\section{Averaged Null Energy Condition in the Present Framework}\label{7.2}
\begin{figure}[!h]
  \centering
  \includegraphics[width=0.45\textwidth]{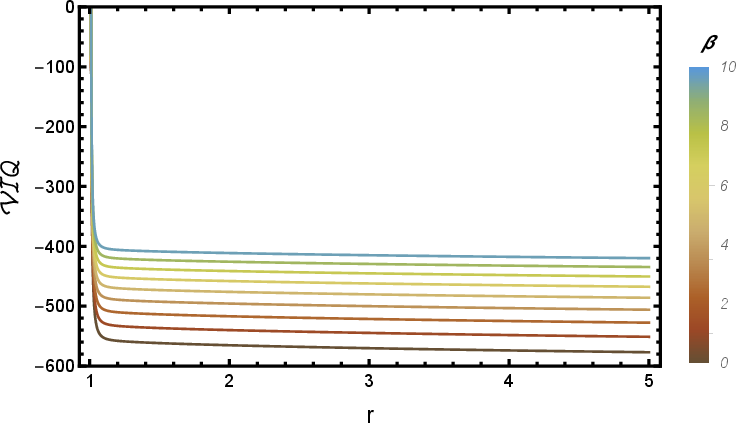}
  \hspace{5mm}
  \includegraphics[width=0.45\textwidth]{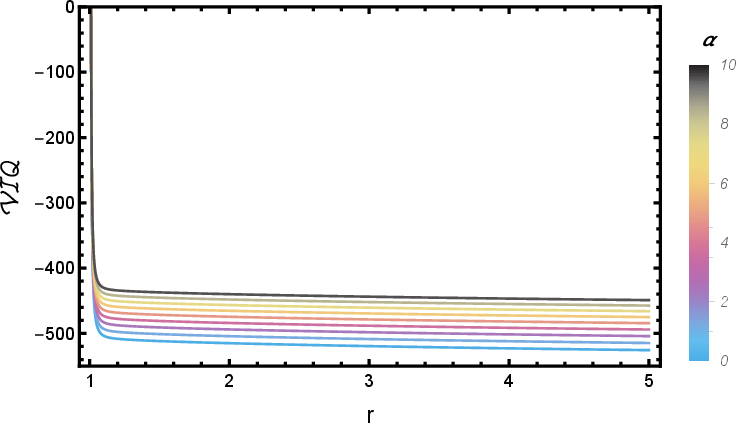}\vspace{5mm}
  \includegraphics[width=0.45\textwidth]{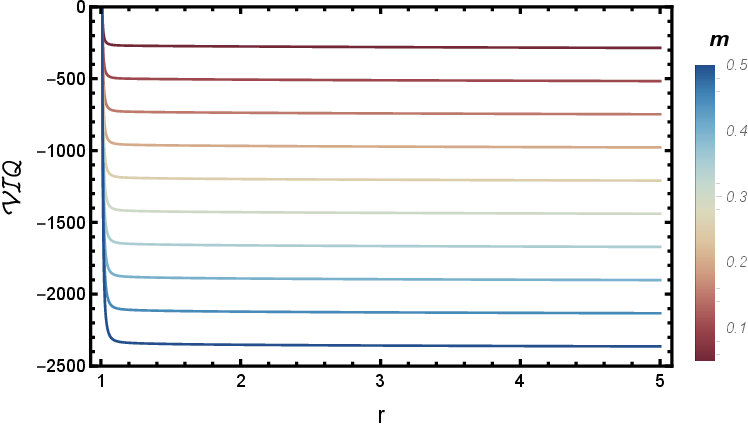}
  \hspace{5mm}
  \includegraphics[width=0.45\textwidth]{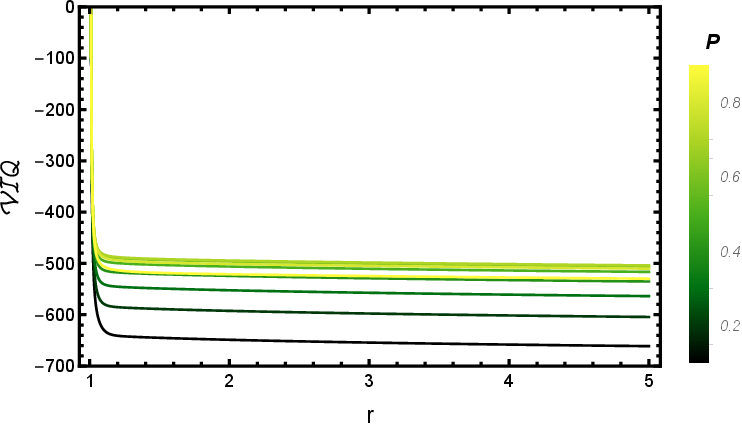}
  \caption{Volume integral quantifier profiles plotted as functions of the radial coordinate $r$, illustrating the effect of varying one parameter at a time while keeping the others fixed at $\mu=-8$ and throat radius $r_t=1$ kpc. Top-left panel: $\beta\in[0,10]~\text{with step size 1.2}$, $\alpha=1$, $m=0.1$, $P=0.5$; top-right panel: $\alpha\in[0,10]~\text{with step size 1.2}$, $\beta=3$, $m=0.1$, $P=0.5$; bottom-left panel: $m\in[0.05,0.5]~\text{with step size 0.05}$, $\alpha=1$, $\beta=3$, $P=0.5$; bottom-right panel: $P\in[0.1,0.9]~\text{with step size 0.1}$, $\alpha=1$, $\beta=3$, $m=0.1$.}\label{anec1}
\end{figure}

In this section, we aim to estimate the total quantity of exotic matter needed to sustain the wormhole by evaluating the violation of ANEC through an integral approach, commonly referred to as the ‘Volume Integral Quantifier’. As noted by Visser et al. \cite{Visser:1995cc}, this quantity is not associated with transverse pressure $p_t$; rather, it is related solely to energy density $\rho$ and radial pressure $p_r$. It is represented in the form of the integral illustrated in the following \cite{Visser:2003yf}.
\begin{equation}
   \mathcal{VIQ}=\oint {(\rho+p_r)d\mathcal{V}}.
 \end{equation}
 with the volume element given by $d\mathcal{V}=r^2 sin\:\theta\: dr \:d\theta \:d\phi$, which may also be expressed as:
  \begin{equation}
     \mathcal{VIQ}= 2\int^\infty_{r_{t}}{(\rho+p_r)\: 4\pi r^2} dr.
 \end{equation}
 The energy-momentum tensor may be excluded for $r_{cut}\geq r_t$, enabling us to express the integral in a modified form as:
  \begin{equation}\label{51}
     \mathcal{VIQ}= 8\pi\int^{r_{cut}}_{r_t}{(\rho+p_r)\:r^2} dr.
 \end{equation}
 With the substitution of $\rho+p_r$ into the previous \cref{51}, the integral $\mathcal{VIQ}$ can be easily evaluated. An important observation is that the integral $\mathcal{VIQ}$ should approach zero as $r_{cut}\to r_t$. Due to the highly complex analytical structure of $\mathcal{VIQ}$, we analyze the behavior of the integral concerning $r$ using a graphical analysis. We now recognize that if the integral $\mathcal{VIQ}$ yields a negative value, it signifies a breach of the ANEC. \cref{anec1} clearly demonstrates that the ANEC is violated near the wormhole throat for all values of the chosen parameters. This violation, interpreted through the Volume Integral Quantifier (VIQ), confirms the necessity of exotic matter in the vicinity of the throat. As a result, the proposed spacetime configurations offer a viable pathway toward realizing physically acceptable, traversable wormhole solutions.
%%%%%%%%%%%%%%%%%%%%%%%%%%%%%%%%%%%%%%%%%%%%%%%%%%%%%%%%%%%%%%%%%%%%%%%%%%%%%%%
%%%%%%%%%%%%%%%%%%%%%%%%%%%%%%%%%%%%%%%%%%%%%%%%%%%%%%%%%%%%%%%%%%%%%%%%%%
%=========================================================================
\section{Thermodynamic analysis of traversable wormholes}\label{sec8}

Wormholes display thermodynamic characteristics akin to those of black holes when examined through local physical parameters. The temperature of black holes and wormholes differs from that of ordinary matter. The form of these objects makes it possible to determine their temperature precisely. The thermal state of these entities can be inferred using just their geometric surface gravity \cite{Hong:2003dj}. If a conventional black hole is applied within GR computations, it becomes clear that a negative temperature is not possible. As the temperature nears absolute zero, black holes exhibit unstable thermodynamic behavior, making the concept of a negative temperature unfeasible \cite{Hong:2003dj}. However, when the same equations are applied to wormholes—accounting for the exotic matter within—they do not exhibit issues with thermodynamic instability, even as the temperature approaches absolute zero. The stabilizing effect of exotic matter in wormholes permits the emergence of negative temperatures \cite{Hong:2003dj}.\\

\begin{figure}[!h]
  \centering
  \includegraphics[width=0.48\textwidth]{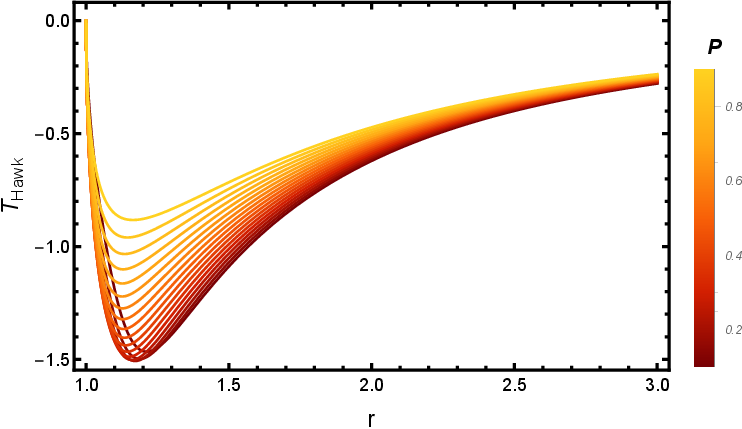}
  \hspace{2mm}
  \includegraphics[width=0.48\textwidth]{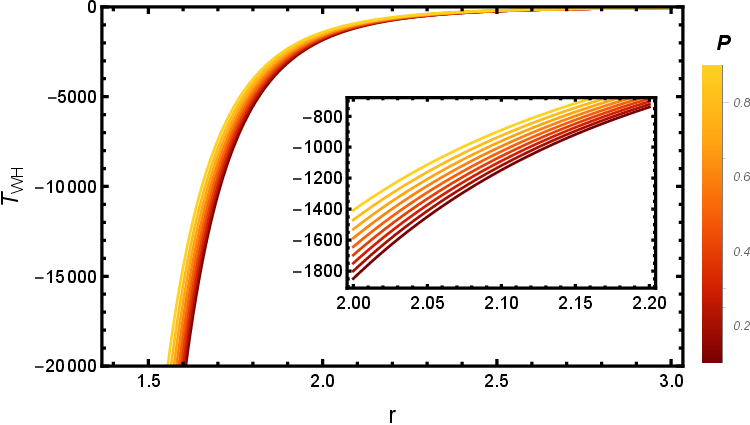}
  \caption{Profiles of Hawking temperature and wormhole temperature as functions of the radial coordinate $r$, for various values of the parameter $P$, with fixed $\mu=-8$ and throat radius $r_t=1$ kpc.}\label{fig8}
\end{figure}

Keathley \cite{keathley2022detectability} provides a description of the Hawking temperature, which is also related to the temperature of wormholes:
\begin{equation}\label{40}
    \begin{aligned}
    T_{Hawk}&~=~\frac{1}{2\pi}\Phi'(r)\sqrt{1-\frac{b(r)}{r}}
            &~=~\frac{\mu  \sqrt{\frac{P  \left(\frac{r^5}{P  r^4+r^4_t (r_t-P ) e^{2 \mu  \left(\frac{1}{r_t}-\frac{1}{r}\right)}}-1\right)}{r}}}{\pi  r^2}.
    \end{aligned}
\end{equation}
As observed in the left panel of \cref{fig8}, the presence of a negative Hawking temperature near the wormhole throat in $\mathcal{F}(Q,\mathcal{L}_{m},\mathcal{T})$ gravity suggests a thermally stable traversable wormhole configuration.\\

By applying the surface gravity to the Hawking temperature expression \cref{40}, allows us to derive the expression for the wormhole temperature \cite{keathley2022detectability}:
\begin{equation}
    \begin{aligned}
    T_{WH}&~=~T_{Hawk}e^{\Phi(r)}&~=~\frac{e^{\Phi(r)}}{2\pi}\Phi'(r)\sqrt{1-\frac{b(r)}{r}}
          &~=~\frac{\mu  e^{-\frac{2 \mu }{r}} \sqrt{\frac{P  \left(\frac{r^5}{P  r^4+r^4_t (r_t-P ) e^{2 \mu  \left(\frac{1}{r_t}-\frac{1}{r}\right)}}-1\right)}{r}}}{\pi  r^2}.
    \end{aligned}
\end{equation}
In a similar manner, the right panel of \cref{fig8} shows that the temperature of the wormhole retains negative values near the throat, reinforcing the indication of thermal stability in the $\mathcal{F}(Q,\mathcal{L}_{m},\mathcal{T})$ gravity scenario.
%%%%%%%%%%%%%%%%%%%%%%%%%%%%%%%%%%%%%%%%%%%%%%%%%%%%%%%%%%%%%%%%%%%%%%%%%%%%%%%
%%%%%%%%%%%%%%%%%%%%%%%%%%%%%%%%%%%%%%%%%%%%%%%%%%%%%%%%%%%%%%%%%%%%%%%%%%%%%%%
\begin{figure}[!h]
  \centering
  \includegraphics[width=0.45\textwidth]{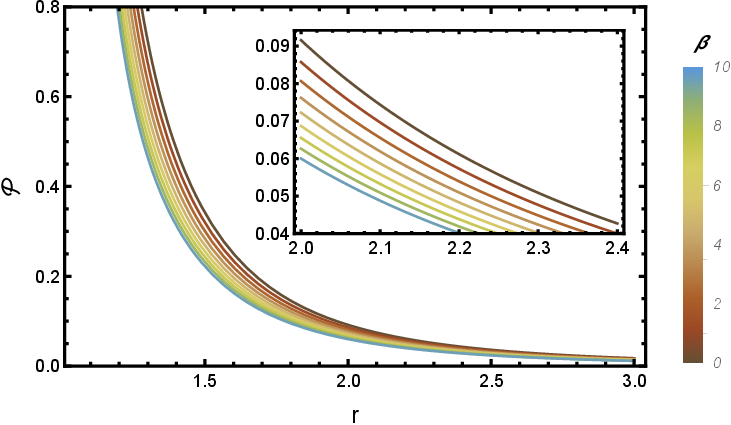}
  \hspace{5mm}
  \includegraphics[width=0.45\textwidth]{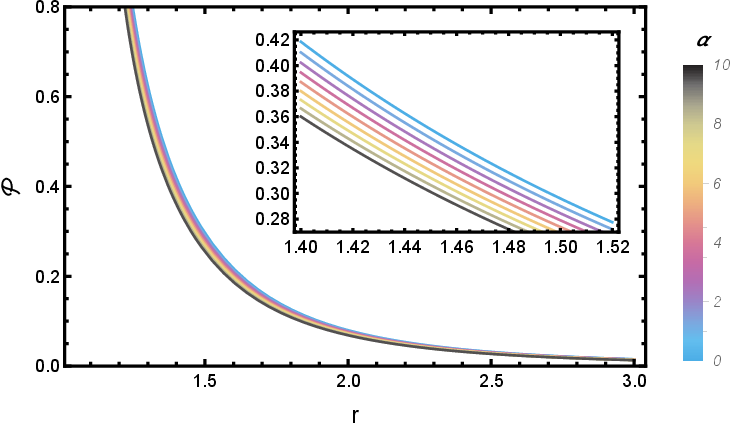}\vspace{5mm}
  \includegraphics[width=0.45\textwidth]{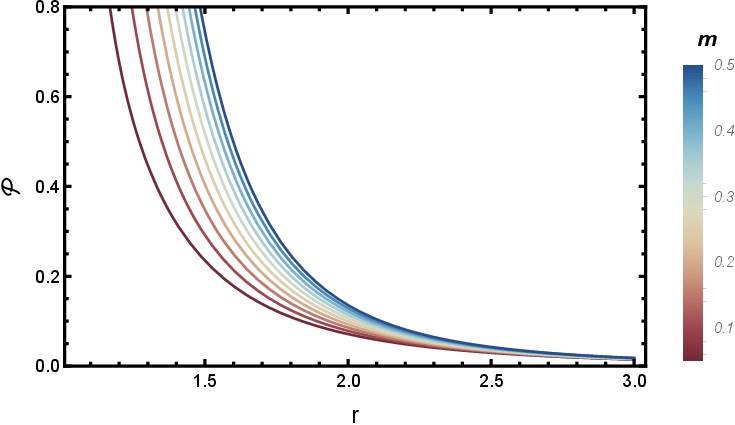}
  \hspace{5mm}
  \includegraphics[width=0.45\textwidth]{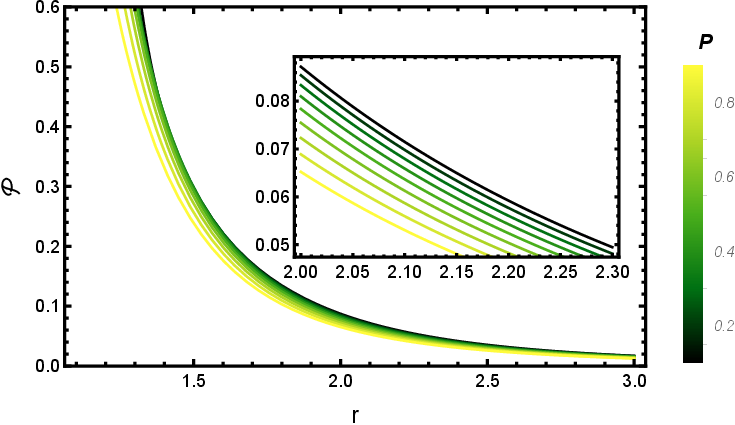}
  \caption{Average pressure profiles plotted as functions of the radial coordinate $r$, illustrating the effect of varying one parameter at a time while keeping the others fixed at $\mu=-8$ and throat radius $r_t=1$ kpc. Top-left panel: $\beta\in[0,10]~\text{with step size 1.2}$, $\alpha=1$, $m=0.1$, $P=0.5$; top-right panel: $\alpha\in[0,10]~\text{with step size 1.2}$, $\beta=3$, $m=0.1$, $P=0.5$; bottom-left panel: $m\in[0.05,0.5]~\text{with step size 0.05}$, $\alpha=1$, $\beta=3$, $P=0.5$; bottom-right panel: $P\in[0.1,0.9]~\text{with step size 0.1}$, $\alpha=1$, $\beta=3$, $m=0.1$.}\label{fig9}
\end{figure}
%%%%%%%%%%%%%%%%%%%%%%%%%%%%%%%%%%%%%%%%%%%%%%%%%%%%%%%%%%%%%%%%%%%%%%%%%%
\begin{figure}[!h]
  \centering
  \includegraphics[width=0.45\textwidth]{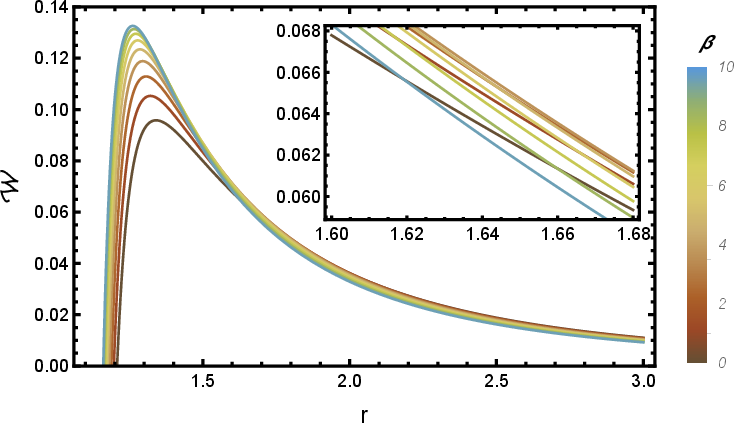}
  \hspace{5mm}
  \includegraphics[width=0.45\textwidth]{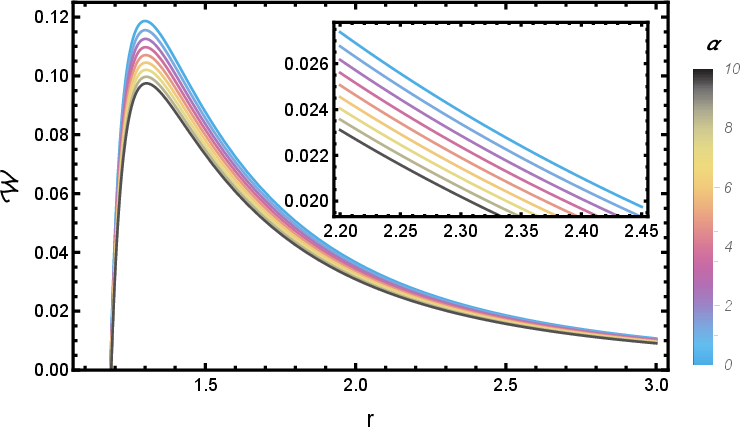}\vspace{5mm}
  \includegraphics[width=0.45\textwidth]{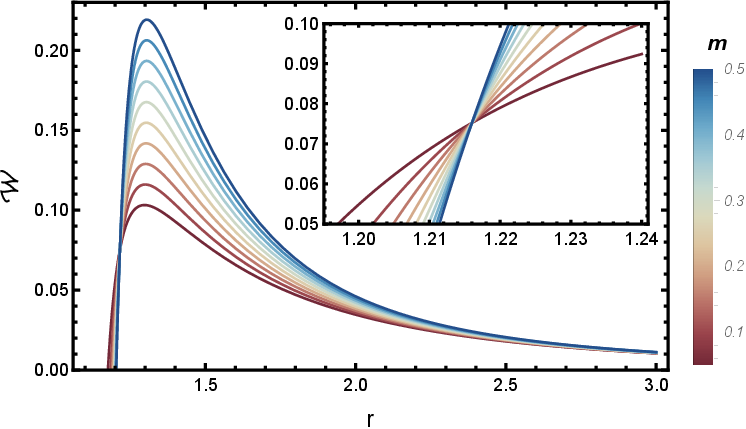}
  \hspace{5mm}
  \includegraphics[width=0.45\textwidth]{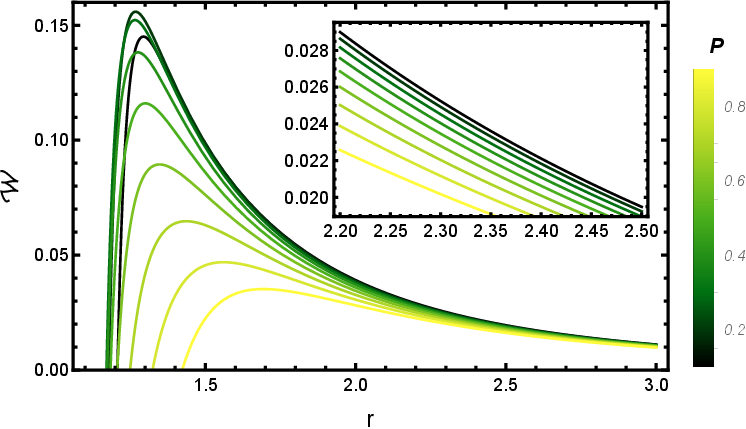}
  \caption{Work density profiles plotted as functions of the radial coordinate $r$, illustrating the effect of varying one parameter at a time while keeping the others fixed at $\mu=-8$ and throat radius $r_t=1$ kpc. Top-left panel: $\beta\in[0,10]~\text{with step size 1.2}$, $\alpha=1$, $m=0.1$, $P=0.5$; top-right panel: $\alpha\in[0,10]~\text{with step size 1.2}$, $\beta=3$, $m=0.1$, $P=0.5$; bottom-left panel: $m\in[0.05,0.5]~\text{with step size 0.05}$, $\alpha=1$, $\beta=3$, $P=0.5$; bottom-right panel: $P\in[0.1,0.9]~\text{with step size 0.1}$, $\alpha=1$, $\beta=3$, $m=0.1$.}\label{fig10}
\end{figure}
%%%%%%%%%%%%%%%%%%%%%%%%%%%%%%%%%%%%%%%%%%%%%%%%%%%%%%%%%%%%%%%%%%%%%
The average pressure within a wormhole is defined as \cite{Callen:450289, Ma:2015llh}
\begin{equation}
    \mathcal{P}=\frac{p_r+2 p_t}{3}.
\end{equation}
For a stable equilibrium configuration, the average pressure should remain positive, as illustrated in Fig. 1 for various selected values of the parameters $\beta$, $\alpha$, $m$, and $P$ across four different plots. Additionally, the temperature is expected to be negative, a concept supported by the work of Hong and Kim \cite{Hong:2003dj}, who proposed that the presence of exotic matter in wormhole models can give rise to such negative temperature distributions.

In a cosmological context, studies have demonstrated that the universal form of the first law of thermodynamics applies to wormholes as follows \cite{Saiedi:2012qk}:
\begin{equation}
    dE=T_{Hawk}\;d\mathcal{S}+\mathcal{W} d\mathcal{V}.
\end{equation}
The entropy of the wormhole is defined as $\mathcal{S}=8\pi r^2$, and its volume is expressed as $\mathcal{V}=\frac{4\pi r^3}{3}$. In this context, $E$ stands for the total energy of the matter, and $\mathcal{W}$ refers to the work density. In this case, the work density associated with the wormhole is defined as follows:
\begin{equation}
    \mathcal{W}=\frac{\rho-p_r}{2}.
\end{equation}
However, by applying the first law of wormhole thermodynamics, the total energy of the configuration can be obtained as follows:
\begin{equation}
    E=\int^r_{r^+_t} {\frac{2}{r} \left(-\pi  r^3  p_r(r)+\pi  r^3 \rho (r)+8 \mu  \sqrt{\frac{P  \left(\frac{r^5}{P  r^4+r^4_t (r_t-P ) e^{2 \mu  \left(\frac{1}{r_t}-\frac{1}{r}\right)}}-1\right)}{r}}\right)}\;\;dr.
\end{equation}

In theory, a wormhole can connect two separate points in spacetime. Introducing negative energy into the wormhole can help stabilize it and prevent it from collapsing. The stability of a traversable wormhole depends on the equilibrium of its stress-energy components. As shown in \cref{fig10}, the work density becomes positive beyond a certain radial distance for all considered parameter sets. A positive work density generates radial tension that helps prevent throat collapse, while a negative total energy signifies the existence of exotic matter required to maintain the wormhole's structure. Together, these factors point to a metastable state that allows temporary traversability without rapid gravitational collapse.
%%%%%%%%%%%%%%%%%%%%%%%%%%%%%%%%%%%%%%%%%%%%%%%%%%%%%%%%%%%%%%%%%%%%
\begin{figure}[!h]
  \centering
  \includegraphics[width=0.45\textwidth]{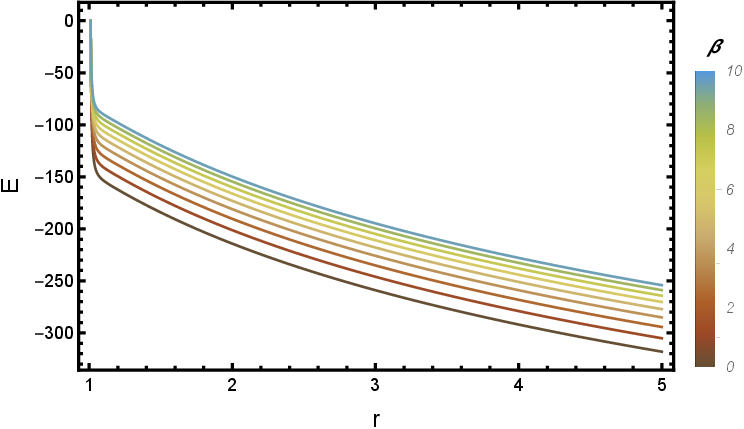}
  \hspace{5mm}
  \includegraphics[width=0.45\textwidth]{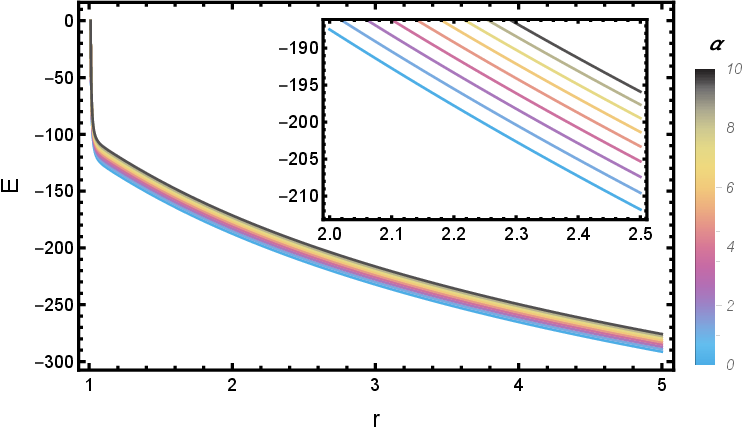}\vspace{5mm}
  \includegraphics[width=0.45\textwidth]{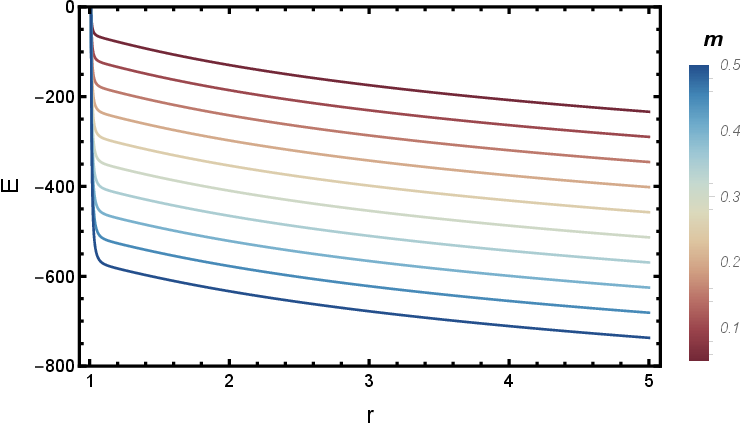}
  \hspace{5mm}
  \includegraphics[width=0.45\textwidth]{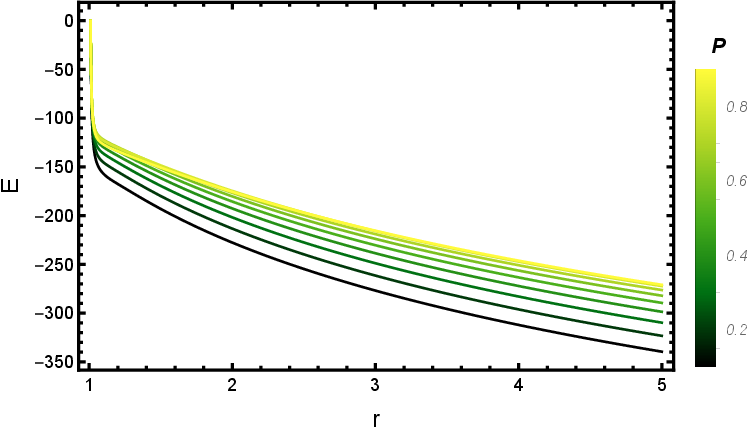}
  \caption{Total energy profiles plotted as functions of the radial coordinate $r$, illustrating the effect of varying one parameter at a time while keeping the others fixed at $\mu=-8$ and throat radius $r_t=1$ kpc. Top-left panel: $\beta\in[0,10]~\text{with step size 1.2}$, $\alpha=1$, $m=0.1$, $P=0.5$; top-right panel: $\alpha\in[0,10]~\text{with step size 1.2}$, $\beta=3$, $m=0.1$, $P=0.5$; bottom-left panel: $m\in[0.05,0.5]~\text{with step size 0.05}$, $\alpha=1$, $\beta=3$, $P=0.5$; bottom-right panel: $P\in[0.1,0.9]~\text{with step size 0.1}$, $\alpha=1$, $\beta=3$, $m=0.1$.}\label{fig10.2}
\end{figure}
%%%%%%%%%%%%%%%%%%%%%%%%%%%%%%%%%%%%%%%%%%%%%%%%%%%%%%%%%%%%%%%%%%%%%%%%%%%%%
 
The expression for the energy flux $\mathcal{F}$ is \cite{martin2011lorentzian}:
\begin{equation}
  \mathcal{F}=\frac{1}{4}e^{\phi(r)}(\rho+p_r)\sqrt{1-\frac{b(r)}{r}}.  
\end{equation}
In wormhole geometries, the condition $\rho+p_r<0$ implies the presence of exotic matter, characterized by an energy flux that draws energy from spacetime rather than contributing to it, as ordinary matter does. This negative energy flux represents a violation of the classical energy conditions, but is essential for maintaining the wormhole structure. However, the magnitude of this flux remains insufficient to reverse the overall gravitational energy variation, thereby preserving the wormhole's stability under small perturbations.
%%%%%%%%%%%%%%%%%%%%%%%%%%%%%%%%%%%%%%%%%%%%%%%%%%%%%%%%%%%%%%%%%%%%
\begin{figure}[htbp]
  \centering
  \includegraphics[width=0.45\textwidth]{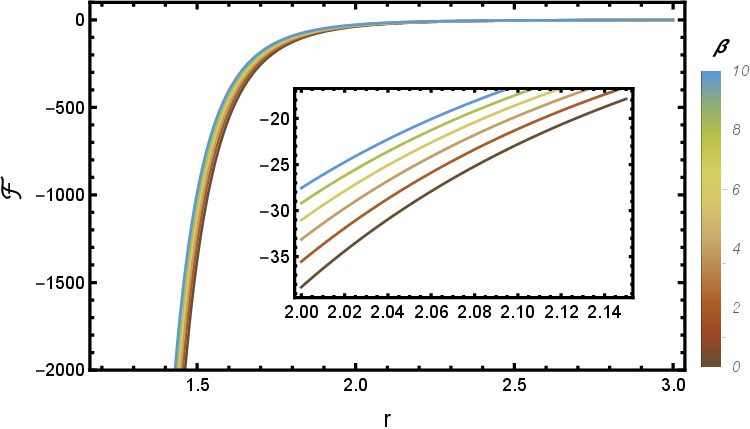}
  \hspace{5mm}
  \includegraphics[width=0.45\textwidth]{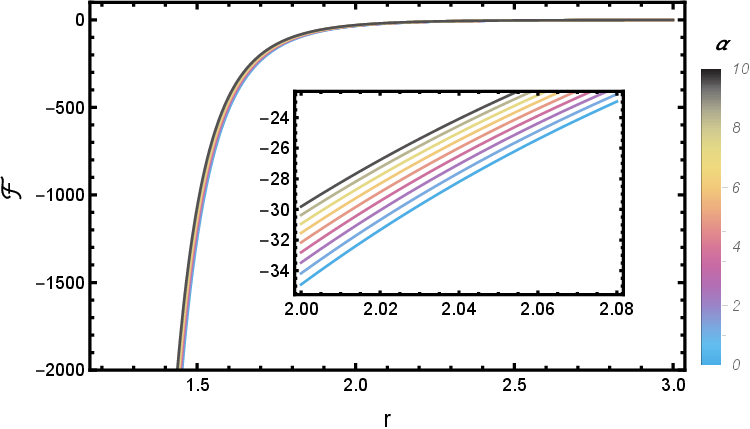}\vspace{5mm}
  \includegraphics[width=0.45\textwidth]{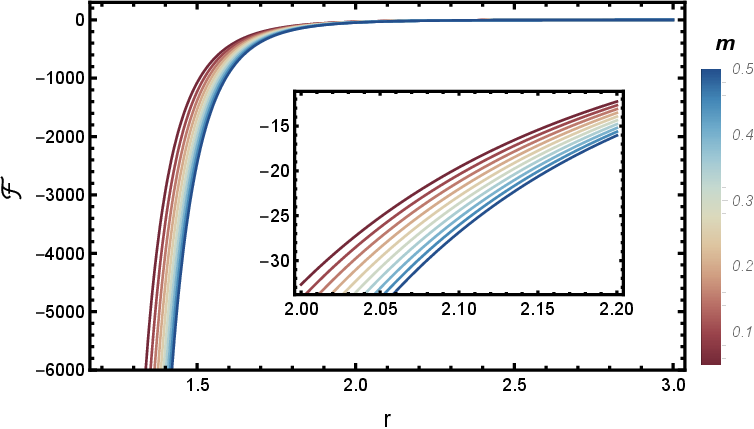}
  \hspace{5mm}
  \includegraphics[width=0.45\textwidth]{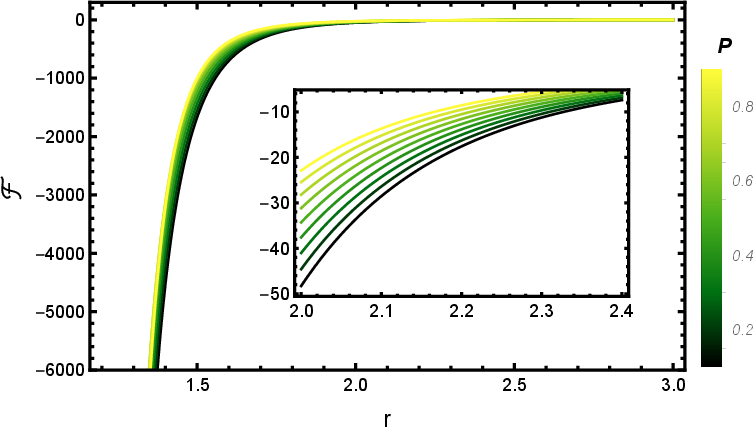}
  \caption{Energy flux profiles plotted as functions of the radial coordinate $r$, illustrating the effect of varying one parameter at a time while keeping the others fixed at $\mu=-8$ and throat radius $r_t=1$ kpc. Top-left panel: $\beta\in[0,10]~\text{with step size 1.2}$, $\alpha=1$, $m=0.1$, $P=0.5$; top-right panel: $\alpha\in[0,10]~\text{with step size 1.2}$, $\beta=3$, $m=0.1$, $P=0.5$; bottom-left panel: $m\in[0.05,0.5]~\text{with step size 0.05}$, $\alpha=1$, $\beta=3$, $P=0.5$; bottom-right panel: $P\in[0.1,0.9]~\text{with step size 0.1}$, $\alpha=1$, $\beta=3$, $m=0.1$.}\label{fig11}
\end{figure}
%%%%%%%%%%%%%%%%%%%%%%%%%%%%%%%%%%%%%%%%%%%%%%%%%%%%%%%%%%%%%%%%%%%%
We now turn to an analysis of the thermodynamic stability of the wormhole configurations. We comply with conventional standards for thermodynamic stability, specifically, $|\frac{\partial\mathcal{P}}{\partial \mathcal{V}}|_T\leq0$ and $C_\mathcal{P}\geq C_\mathcal{V}\geq0$. Here, $\mathcal{P}=\frac{p_r+2p_t}{3}$ denotes the average pressure, and $C_\mathcal{P}$ and $C_\mathcal{V}$ correspond to the specific heats at constant pressure and constant volume, respectively. During this wormhole exploration, we focus solely on calculating the specific heat $C_\mathcal{V}$, without accounting for the effects of pressure:
\begin{equation}
    C_\mathcal{V}=T_{Hawk}\frac{d\mathcal{S}}{dT_{Hawk}}=-\frac{32 \pi  r^2 \left(r^5-P  r^4-r^4_t (r_t-P ) e^{2 \mu  \left(\frac{1}{r_t}-\frac{1}{r}\right)}\right) \left(P  r^4+r^4_t (r_t-P ) e^{2 \mu  \left(\frac{1}{r_t}-\frac{1}{r}\right)}\right)}{P  r^8 (4 r-5 P )+2 r^4 r^4_t (\mu -5 P ) (r_t-P ) e^{2 \mu  \left(\frac{1}{r_t}-\frac{1}{r}\right)}-5 r^8_t (r_t-P )^2 e^{4 \mu  \left(\frac{1}{r_t}-\frac{1}{r}\right)}}
\end{equation}
%%%%%%%%%%%%%%%%%%%%%%%%%%%%%%%%%%%%%%%%%%%%%%%%%
\begin{figure}[htbp]
  \centering
  \includegraphics[width=0.50\textwidth]{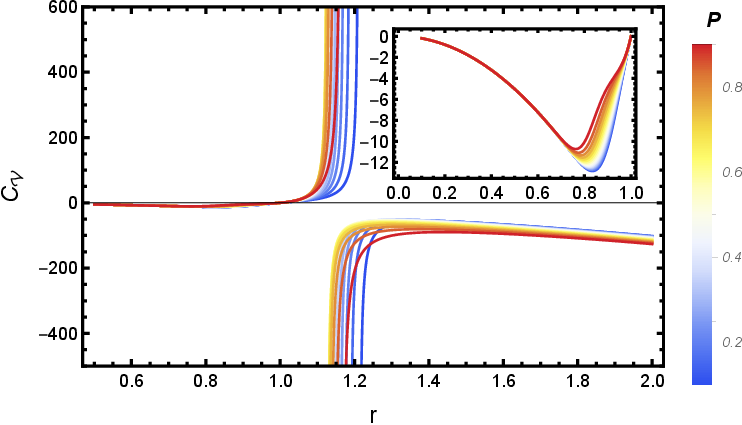}
  \caption{Variation of specific heat $C_\mathcal{V}$ with radial distance $r$ for various values of parameter $P\in[0.1,0.9]~\text{with step size 0.1}$, keeping $\mu=-8$ and throat radius $r_t=1$ kpc fixed.}\label{fig12}
\end{figure}
%%%%%%%%%%%%%%%%%%%%%%%%%%%%%%%%%%%%%%%%%%%%%%%%%%%%%%%%%%%%
\cref{fig12} depicts the phase transition characteristics of the specific heat. It can be observed that for $r<r_t=1$, the wormhole configuration exhibits thermodynamic instability. At $r=r_t$, a transition to a stable phase occurs, marked by the specific heat becoming positive ($C_\mathcal{V}>0$). However, beyond $r>1.2$, the system undergoes a second transition, returning to a globally unstable state. 
%%%%%%%%%%%%%%%%%%%%%%%%%%%%%%%%%%%%%%%%%%%%%%%%%%%%%%%%%%%%%%%%%%%
\begin{table} 
\centering
\caption{Specific heat \( C_\mathcal{V} \) and corresponding stability condition for various values of parameter \( P \)}
\renewcommand{\arraystretch}{1.4}
\begin{xltabular}
{\linewidth}{>{\scriptsize}p{0.05\textwidth} >{\scriptsize}p{0.07\textwidth} >{\scriptsize}p{0.07\textwidth} >{\scriptsize}p{0.07\textwidth} >{\scriptsize}p{0.07\textwidth} >{\scriptsize}p{0.07\textwidth} >{\scriptsize}p{0.07\textwidth} >{\scriptsize}p{0.07\textwidth}>{\scriptsize}p{0.07\textwidth}>{\scriptsize}p{0.07\textwidth}>{\scriptsize}p{0.07\textwidth}>{\scriptsize}p{0.07\textwidth}}

\hline\hline
\multirow{2}{*}{\textbf{\large\( r \)}} & \multicolumn{2}{c}{\textbf{\( P = 0.1 \)}} & \multicolumn{2}{c}{\textbf{\( P = 0.3 \)}} & \multicolumn{2}{c}{\textbf{\( P = 0.5 \)}} & \multicolumn{2}{c}{\textbf{\( P = 0.7 \)}} & \multicolumn{2}{c}{\textbf{\( P = 0.9 \)}} \\
\cline{2-11}
 & \( C_\mathcal{V} \) & Stability & \( C_\mathcal{V} \) & Stability & \( C_\mathcal{V} \) & Stability & \( C_\mathcal{V} \) & Stability & \( C_\mathcal{V} \) & Stability \\
\hline\hline
0.8 & -12.4507 & Unstable   & -12.3378 & Unstable   & -12.1412 & Stable   & -11.7128 & Unstable & -10.055 & Unstable \\
1 & 0 & Marginal   & 0 & Marginal   & 0 & Marginal   & 0 & Marginal & 0 & Marginal \\
1.1 & 11.7837 & Stable   & 26.2413 & Stable   & 48.5038 & Stable & 66.4753 & Stable & 41.851 & Stable \\
1.12 & 16.6946 & Stable   & 53.7121 & Stable & 173.356 & Stable & 325.82 & Stable & 78.649 & Stable \\
1.13 & 20.1622 & Stable & 87.4309 & Stable & 1714.64 & Stable & -984.373 & Unstable & 113.911 & Stable \\
1.14 & 24.7088 & Stable & 180.513 & Stable & -282.305 & Unstable & -236.452 & Unstable & 178.751 & Stable \\
1.15 & 30.8789 & Stable & 1444.84 & Stable & -146.367 & Unstable & -147.732 & Unstable & 334.463 & Stable \\
1.16 & 39.6325 & Stable & -303.898 & Unstable & -105.732 & Unstable & -113.853 & Unstable & 1191.44 & Stable \\
1.17 & 52.8405 & Stable & -152.476 & Unstable & -86.5604 & Unstable & -96.3037 & Unstable & -1051.12 & Unstable \\
1.21 & 1478.42 & Stable & -68.3379 & Unstable & -60.9523 & Unstable & -70.9228 & Unstable & -174.876 & Unstable \\
1.22 & -404.776 & Unstable & -63.3322 & Unstable & -58.6867 & Unstable & -68.5198 & Unstable & -153.696 & Unstable \\
1.25 & -106.024 & Unstable & -55.5323 & Unstable & -55.1238 & Unstable & -64.5721 & Unstable & -120.847 & Unstable \\
\hline
\textbf{Stable \( r \)-range} 
& \multicolumn{2}{c}{\( (1, 1.21] \)} 
& \multicolumn{2}{c}{\( (1, 1.15] \)} 
& \multicolumn{2}{c}{\( (1, 1.13] \)} 
& \multicolumn{2}{c}{\( (1, 1.12] \)} 
& \multicolumn{2}{c}{\( (1, 1.16] \)} \\
\hline
\end{xltabular}
\label{table2}
\end{table}
%%%%%%%%%%%%%%%%%%%%%%%%%%%%%%%%%%%%%%%%%%%%%%%%%%%%%%%%%%%%%%%%
To supplement this, \cref{table2} presents the specific intervals of thermal stability (where $C_\mathcal{V} > 0$) for different values of the parameter $P$. The identified stable radial ranges are $(1, 1.21]$ for $P = 0.1$, $(1, 1.15]$ for $P = 0.3$, $(1, 1.13]$ for $P = 0.5$, $(1, 1.12]$ for $P = 0.7$, and $(1, 1.16]$ for $P = 0.9$. Although no strict monotonic trend is observed, the stability region remains consistently confined to a narrow band just outside the throat for all parameter values. This confirms that the equilibrium configurations in $\mathcal{F}(Q,\mathcal{L}_{m},\mathcal{T})$ gravity are highly localized and sensitive to the choice of $P$, with specific heat acting as a key thermodynamic indicator of wormhole stability.
%===================================================================================================

\newpage
\section{Concluding Remarks}\label{sec10}
The present work is dedicated to the study of traversable wormholes in the realm of $\mathcal{F}(Q, \mathcal{L}_{m}, \mathcal{T})$ gravitational theory, along with an analysis of their associated thermodynamical properties. We began by formulating the mathematical structure of the theory and deriving the altered gravitational field equations corresponding to static and spherically symmetric spacetime geometry. Using the embedding class-1 approach in conjunction with Karmarkar’s condition, we obtained an exact wormhole shape function. The shape function was then analyzed and confirmed to satisfy the essential traversability criteria. We have constructed and presented visual representations of the spacetime embedding in two and three dimensions based on the derived shape function, using selected values of the parameter $P$. 
Furthermore, the embedding profile $z(r)$ and the proper radial function $l(r)$ have been numerically evaluated and tabulated for specific radial positions r ($>r^+_t$), where the throat radius is taken as $r_t=1$ kpc. Subsequently, we analyzed the behavior of standard energy constraints, namely NEC, WEC, SEC, and DEC, in close proximity to the throat of the wormhole, focusing on the domain $r>r^+_t$. Following that, the ANEC was examined using the VIQ to evaluate the overall violation throughout the wormhole spacetime. Afterward, a thermodynamic analysis of the wormhole was carried out to explore its thermal properties under the theoretical setting of $\mathcal{F}(Q,\mathcal{L}_{m},\mathcal{T})$ gravity. In our investigation, exotic matter is assumed to be indispensable for sustaining the constructed wormhole solutions. To compute the overall energy associated with the wormhole geometry, we adopted a generalized thermodynamic approach that integrates an extended first law and the concept of generalized surface gravity. To assess the thermodynamic stability of the wormhole configuration, we calculated the specific heat without incorporating the effects of pressure, providing insight into the system's thermal response under energy fluctuations. The core results obtained in this study are systematically presented as follows.\\

As depicted in the graphical results shown in \cref{fig1}, it is evident that the shape function $b(r)$, as defined in \cref{36}, meets all the fundamental criteria required for a physically viable traversable wormhole. Specifically, it meets the requirements for throat condition, flaring-out criterion, and asymptotic flatness, as described in conditions ($\ref{cond2}$) through (\ref{cond5}). These properties hold for the selected parameter choices $P$, thereby confirming the theoretical consistency of the proposed wormhole geometry. Furthermore, \cref{fig3.1} offers a clear geometric interpretation of the wormhole structure through its two- and three-dimensional visualizations of the embedded geometry, enhancing our understanding of the spatial curvature. In addition, the variation of the proper radial distance depending on the radial coordinate, as illustrated in \cref{fig3.2}, adds credibility to the physical acceptability of the wormhole configuration. Furthermore, the data presented in \cref{table1} reveal that $l(r)$ consistently exceeds the embedding surface function $z(r)$ for all $r>r_t$. This inequality satisfies a key geometrical condition required for the plausibility of constructing a physically acceptable traversable wormhole, thereby reinforcing the validity of the proposed model. Therefore, the shape function $b(r)$, derived using the Karmarkar condition \cref{kar} for the chosen redshift function $\Phi(r)$ as given in \cref{33}, proves to be a viable solution for describing the geometry of a traversable wormhole.\\

The energy conditions were thoroughly examined for four different parameter sets, specifically varying $\beta$, $\alpha$, $m$, and $P$, as depicted in \crefrange{fig4.1}{fig4.4}.  From the detailed analysis of these plots, it is clear that the NEC, among others, is violated near the wormhole throat in all cases. Such violations are a hallmark of exotic matter, which is a known requirement for sustaining traversable wormholes according to the principles of GR. Remarkably, our results show that this characteristic persists even as part of the study of modified gravitational frameworks like $\mathcal{F}(Q,\mathcal{L}_{m},\mathcal{T})$ gravity. Therefore, we conclude that the existence of exotic matter remains an essential feature for maintaining traversable wormholes in this theoretical setting as well.\\

To measure the quantity of exotic matter required to sustain a wormhole, we evaluated the ANEC near the throat region. This analysis, presented in \cref{anec1} for different values of the parameters, reveals that the ANEC integrals produce consistently negative values in all cases. Such results clearly indicate the violation of the ANEC, reinforcing the necessity of exotic matter for the stability of wormhole structures. Consequently, even within the context of $\mathcal{F}(Q,\mathcal{L}_{m},\mathcal{T})$ gravity, a significant presence of exotic matter is essential to support traversable wormhole configurations.\\

Several key insights have emerged based on our investigation of the wormhole's thermodynamic properties. As illustrated in \cref{fig8}, both the wormhole temperature (right panel) and the Hawking temperature (left panel) exhibit negative behavior, which, when considered along with the positive pressure profile shown in \cref{fig9}, supports the existence of a stable thermodynamic equilibrium within the exotic matter regime. Further strengthening this stability, the system demonstrates positive work density (\cref{fig10}) and a negative total energy (\cref{fig10.2}), indicating resistance to gravitational collapse and favoring long-term equilibrium. The negative energy flux observed in \cref{fig11} reflects an outward flow of energy from the throat, a signature characteristic of exotic matter, and an essential feature for sustaining traversable wormholes. Moreover, the specific heat remains positive near the throat, as shown in \cref{fig12}, indicating local thermal stability. The specific heat profile also reveals a transition between stable and unstable regions, suggesting the presence of phase-like behavior in the thermodynamic structure of the wormhole. Collectively, these findings highlight the intricate interplay between geometry, matter content, and thermodynamic conditions in supporting a stable wormhole configuration. Our results emphasize that even within modified gravity frameworks, such as $\mathcal{F}(Q,\mathcal{L}_{m},\mathcal{T})$ gravity, exotic matter plays a crucial role not only in maintaining the geometric viability of wormholes but also in ensuring their thermodynamic and energetic stability.\\

Compared with recent work such as \cite{Ditta:2025wwx}, which investigates wormhole thermodynamics in the framework of extended teleparallel gravity, we find that our thermodynamic behavior exhibits strong similarities. In both studies, the wormhole and Hawking temperatures remain negative, the average pressure is positive, and the total energy and energy flux are negative, each indicating thermodynamic equilibrium sustained by exotic matter. However, a notable distinction lies in the work density: while it is negative in \cite{Ditta:2025wwx}, suggesting a repulsive or tension-like internal response, our solution yields a positive work density, which plays a crucial role in counteracting the gravitational pull and contributes to the mechanical stability of the wormhole structure. This highlights a key difference in the equilibrium structure of wormholes in $\mathcal{F}(Q,\mathcal{L}_{m},\mathcal{T})$ gravity compared to its teleparallel counterpart.\\
 
To conclude, our investigation in the context of $\mathcal{F}(Q,\mathcal{L}_{m},\mathcal{T})$ gravity confirms the existence of stable traversable wormhole solutions supported by a suitable shape function derived via the Karmarkar condition. Despite satisfying all geometric and thermodynamic criteria, the violation of energy conditions such as NEC and ANEC indicates the unavoidable presence of exotic matter. Remarkably, we have emphasized the influence of the extended STG, i.e. $\mathcal{F}(Q, \mathcal{L}_{m}, \mathcal{T})$ gravity, throughout the entire discussion by providing detailed graphical illustrations for a selected range of model parameters $m$, $\alpha$, and $\beta$. Overall, our findings emphasize that even in $\mathcal{F}(Q,\mathcal{L}_{m},\mathcal{T})$ gravity, exotic matter remains essential for maintaining wormhole stability, paving the way for future studies of dynamic behavior and observational prospects.\\\\

\section*{\normalsize\bf{ DATA AVAILABILITY}}
This research is purely theoretical, with all findings derived from analytical equations. No new data were generated during the course of this study; therefore, the article does not include any associated data sets.

\section*{\normalsize\bf{ACKNOWLEDGEMENTS}}
RM is thankful to UGC, Govt. of India,
for providing a Senior Research Fellowship (NTA Ref. No.: 221610058422).

 \printbibliography

\end{document}